\documentclass[aps,prd,twocolumn,showpacs,superscriptaddress,nofootinbib,10pt]{revtex4-2}

\usepackage[dvipsnames]{xcolor}
\usepackage{multirow}

\usepackage{amsmath,amssymb,amsfonts}
\usepackage{amsthm}      % for theorems
\usepackage{cases}       % for piecewise functions

\usepackage[dvipsnames]{xcolor}
\usepackage{graphicx}
\usepackage[T1]{fontenc}

\usepackage{hyperref}
\usepackage[nameinlink]{cleveref}

\begin{document}

%%%%%  TITLE  
\title[Maximum entropy reconstruction of the black hole greybody factors from its Korteweg-de Vries integrals]{
Maximum entropy reconstruction of the black hole greybody factors from its Korteweg-de Vries integrals}

%%%%% AUTHORS    
\author{Noah Alberca}
\email{noah.alberca@uab.cat}
\affiliation{Departament de F\'{\i}sica, Universitat Aut\`onoma de Barcelona, E-08193 Bellaterra, Spain}
\affiliation{Institut de Ci\`encies de l'Espai (ICE, CSIC), Campus UAB, Carrer de Can Magrans s/n, 08193 Cerdanyola del Vall\`es, Spain}
	
\author{Michele Lenzi}
\email{michele.lenzi@ulb.be}
\affiliation{Physique Théorique et Mathématique, Université Libre de Bruxelles (ULB), Campus Plaine, Building NO, CP 231, Boulevard du Triomphe B-1050, Bruxelles, Belgique}
	
\author{Carlos F. Sopuerta}
\email{carlos.f.sopuerta@csic.es}
\affiliation{Institut de Ci\`encies de l'Espai (ICE, CSIC), Campus UAB, Carrer de Can Magrans s/n, 08193 Cerdanyola del Vall\`es, Spain}
\affiliation{Institut d'Estudis Espacials de Catalunya (IEEC), Edifici RDIT, C/ Esteve~Terradas, 1, desp.~212, Castelldefels~08860, Spain}

%%%%% ABSTRACT
\begin{abstract}
The dynamics of perturbed non-rotating black holes (BHs) can be described by master wave equations with a potential barrier. Recent work has shown that these equations possess an infinite number of symmetries generated by the Korteweg-de Vries (KdV) hierarchy, leading to conserved quantities (KdV integrals) that fully determine the BH greybody factors via a moment problem. In this paper, we introduce and validate a new method to solve this moment problem using the maximum entropy principle. The method reconstructs, at each order of truncation, the least-biased transmission probability compatible with a finite set of KdV integrals, guaranteeing positivity by construction. We first validate the technique against the exactly solvable P\"oschl-Teller potential. We then apply it to the Regge-Wheeler potential governing axial perturbations of a Schwarzschild BH, where no exact solution is available; the resulting reconstruction agrees closely with the fourth-order Wentzel-Kramers-Brillouin (WKB) approximation across several multipoles. This is the first application of the maximum entropy method to the reconstruction of BH greybody factors, demonstrating that a modest number of KdV integrals is sufficient to recover them accurately.
\end{abstract}
%%%%%

%%%%% TITLE
\maketitle

%%%%% INTRODUCTION	
\section{Introduction} \label{sec:introduction}
Black holes (BHs)~\cite{Ruffini:1971bza}  are one of the most radical predictions of General Relativity~\cite{Hawking:1973uf,Wald:1984cw,Novikov:1989sz,Chandrasekhar:1992bo}. 
Decades of theoretical and observational work have established BHs as central objects in astrophysics, cosmology, and fundamental physics. 
In general relativity (GR), two conjectures hold up the standard picture of how BHs form and what they look like. 
Gravitational collapse generically produces singularities~\cite{Penrose:1964wq} and the \emph{weak cosmic censorship} conjecture states that, for generic initial data, such singularities are hidden behind event horizons~\cite{Penrose:1969pc,Wald:1974np}. 
On the other hand, the \emph{no-hair} conjecture~\cite{Bekenstein:1996pn,Chrusciel:2012jk} states that the resulting stationary BH is characterized by just three parameters: mass, intrinsic angular momentum (spin), and electric charge. That is, it belongs to the Kerr-Newman family~\cite{Kerr:1963ud,Newman:1965my}. The latter is supported by the uniqueness theorems for stationary, asymptotically flat electrovacuum BHs~\cite{Israel:1967wq,Israel:1967za,Carter:1971zc,Hawking:1971vc,Robinson:1975bv,Mazur:1982db,Chrusciel:2012jk}, by the decay of perturbations in collapse~\cite{Price:1971fb,Price:1972pw}, and by the nonlinear stability of slowly rotating Kerr BHs~\cite{Klainerman:2021qzy,Giorgi:2022omp}, but it does not extend to arbitrary matter content~\cite{Herdeiro:2015waa,Chrusciel:2012jk}. The charge is generally regarded as negligible for astrophysical BHs: any significant net charge is expected to be rapidly neutralized by selective accretion from the surrounding plasma~\cite{Eardley:1975kp,Zajacek:2019kla}, and a rotating BH immersed in a magnetic field only acquires the tiny electric charge~\cite{Wald:1974np}.

Observations of BH candidates are accumulating from various sources, including X-rays~\cite{2016A&A...587A..61C} and gravitational waves~\cite{LIGOScientific:2018mvr,LIGOScientific:2021usb,KAGRA:2021vkt}, with future detectors promising many more candidates with improved sensitivities~\cite{Sathyaprakash:2012jk,Evans:2021gyd,LISA:2017pwj}.
There are physical phenomena in the cosmos involving BHs, like scattering processes of different matter fields and quasinormal mode (QNM) oscillations, which are of particular relevance. Scattering theory for BHs~\cite{Futterman:1988ni,Andersson:2000tf} provides a powerful tool to investigate the nature of the scatterer by collecting asymptotic data. When the scattered particles or waves only induce small deviations from the BH background geometry, BH perturbation theory (BHPT) is the appropriate framework to study them. As found in pioneering works~\cite{Regge:1957td,Cunningham:1978cp,Cunningham:1979px,Cunningham:1980cp,Zerilli:1970la,Zerilli:1970se,Moncrief:1974vm}, and generalized to higher dimensions~\cite{Ishibashi:2003ap,Kodama:2003kk}, it turns out that the perturbations can be described by a set of master wave-like equations with a potential barrier. Actually, all the physical observables can be obtained from the solution, the master wave functions. On the other hand, in the frequency domain, the master equations become time-independent Schr\"odinger equations, where the energy would correspond to minus square of the frequency of the corresponding mode.
	
In this context, the \emph{greybody factors} are key physical quantities as they represent the transmission (and reflection) probabilities associated with the BH potential barrier (the S-matrix for BH scattering), encoding deviations from the black-body spectrum caused by the interaction with the background potential~\cite{Hawking:1975vcx}. The poles of the analytic extension of the transmission coefficient correspond to the BH QNMs~\cite{Kokkotas:1999bd,Nollert:1999re,Berti:2009kk,Konoplya:2011qq}, which are crucial in gravitational wave astronomy as they may contain the fingerprints of new physics~\cite{Barack:2018yly,Perkins:2020tra,Barausse:2020rsu,LISA:2022kgy,ET:2025xjr}.
	
There is an extensive literature  on calculations of greybody factors, including low-frequency~\cite{Page:1976df,Unruh:1976fm,Starobinsky:1973aij,Starobinskil:1974nkd} and high-frequency~\cite{Sanchez:1976fcl,Sanchez:1976xm,Sanchez:1977vz} limits, different Wentzel–Kramers–Brillouin (WKB) approximations~\cite{Schutz:1985km,Iyer:1986np,Iyer:1986nq,Konoplya:2003ii,Matyjasek:2017psv,Konoplya:2019hlu}, bounding of the Bogoliubov coefficients~\cite{Visser:1998ke,Boonserm:2008zg}, and monodromy methods~\cite{Neitzke:2003mz,Castro:2013lba,Harmark:2007jy}.
	
Recent work~\cite{Lenzi:2021wpc,Lenzi:2021njy,Lenzi:2022wjv,Lenzi:2024tgk} has revealed a rich structure in the space of master equations and functions describing Schwarzschild BH perturbations. All possible master equations are connected by Darboux transformations (DTs) (see also~\cite{Lenzi:2025man,Lenzi:2026aay}), establishing their physical equivalence. This leads, in particular, to the isospectrality of the QNM spectrum, in the sense that even- and odd-parity modes share the same spectrum. Additionally, these equations admit an infinite number of symmetries generated by the Korteweg-de Vries (KdV) hierarchy, connecting in this way with the world of integrable systems~\cite{Jaramillo:2024qjz}, which lead to an infinite number of conserved quantities (KdV integrals) that are invariant under DTs~\cite{Lenzi:2021njy}. Remarkably, these KdV integrals fully determine the BH greybody factors via a \emph{moment problem}~\cite{Lenzi:2022wjv,Lenzi:2023inn}. The KdV integrals turn out to be, up to a rescaling, the moments of a distribution function built out of the logarithm of the transmission probability.  In~\cite{Lenzi:2023inn}, a method to invert the BH moment problem to recover the transmission coefficient in terms of the KdV integrals was introduced. The method uses Pad\'e approximants and the Laplace transform. Another application of the KdV integrals is the study of stability properties of the greybody factors and QNM spectrum~\cite{Lenzi:2025kqs}.
	
~
	
\noindent{\em Executive Summary}. In this paper, we develop and validate a new method to solve the BH moment problem using the maximum entropy principle~\cite{Meadpapanicolau1984,Shannon1948,Jaynes1957a,Jaynes1957b}. The method reconstructs, at each order of truncation, the least-biased probability distribution compatible with a given set of moments, guaranteeing positivity at every order. We first validate the technique against the exactly solvable P\"oschl-Teller potential~\cite{Poschl:1933zz}, where we can compare with the exact greybody factor. We then apply it to the Regge-Wheeler potential~\cite{Regge:1957td} governing axial perturbations of a Schwarzschild BH, comparing our results with the WKB approximation.  We use two different approaches, one based on integration on the original infinite frequency domain and, another one based on a compactification method that transform the integrals on an infinite domain to a finite one, improving in this way the performance of the method.
This is the first application of the maximum entropy method to the reconstruction of BH greybody factors, demonstrating that a modest number of KdV integrals is sufficient to recover them accurately.
	
~
	
\noindent{\em Structure of the Paper}. 
In Section~\ref{sec:bhs} we review BHPT for the Schwarzschild case, introducing the master equations and the role of DTs. 
In Section~\ref{sec:kdv} we review the KdV symmetries and the moment problem formulation. In Section~\ref{sec:solving-BH-moment-problem} we summarize the mathematical aspects of the BH moment problem, recall the Pad\'e-based inversion of Ref.~\cite{Lenzi:2023inn}, and introduce the dimensionless variables used in the rest of the paper.
In Section~\ref{sec:maxent} we introduce the maximum entropy method and its numerical implementation. 
In Section~\ref{sec:pt} we validate the method against the P\"oschl-Teller potential and identify the limitations of the implementation. 
In Section~\ref{sec:compact} we address these limitations by reformulating the problem on a compact interval, as a Hausdorff moment problem, and compare both formulations using the P\"oschl-Teller potential. 
In Section~\ref{sec:rw} we apply the method to the Regge-Wheeler potential, choosing the formulation and the number of moments according to the multipole $\ell$. 
We conclude in Section~\ref{sec:conclusions} with a discussion of our results and future directions.
	
~
	
Throughout this paper, unless otherwise stated, we use geometric units with $G = c = 1$.

%%%%% BH PERTURBATION THEORY AND MASTER EQUATIONS 
\section{BH Perturbation Theory and Master Equations} \label{sec:bhs}
The study of perturbations of non-rotating BHs started with the seminal work of Regge and Wheeler~\cite{Regge:1957td}. The background spacetime is described by the Schwarzschild metric~\cite{Schwarzschild:1916uq}:
\begin{equation}
d\hat{s}{}^2 = \hat{g}^{}_{\mu\nu}dx^\mu dx^\nu = -f(r)\,dt^2+\frac{dr^2}{f(r)}+r^2d\Omega^2\,,
\label{schwarzschild-metric}
\end{equation}
where we use a hat to denote quantities defined in the Schwarzschild background spacetime. The metric is completely determined by the function 
\begin{equation}
f(r) = 1 - \frac{r^{}_s}{r} \,,   
\end{equation}
with $r_s = 2GM/c^2 = 2M$ being the Schwarzschild radius, and $d\Omega^2$ denotes the line element of the 2-sphere ($S^{2}$). In standard spherical coordinates its line element is: $d\Omega^2 = d\theta^2 + \sin^2\theta\,d\phi^2\,$.
	
We can introduce metric perturbations of the Schwarzschild spacetime, $h_{\mu\nu}$, satisfying $|h_{\mu\nu}| \ll |\hat{g}_{\mu\nu}|$, via
\begin{equation}
g^{}_{\mu\nu} = \hat{g}^{}_{\mu\nu} + h^{}_{\mu\nu}\,.   
\end{equation}
Taking into account the spherical symmetry of the background, we can expand the perturbations in terms of scalar, vector and tensor spherical harmonics, separating in this way the dependence on $(t,r)$ from the angular dependence. Moreover, it is well-known that the perturbative Einstein equations decouple for each harmonic $(\ell,m)$ and parity\footnote{The spherical harmonics can be divided according to how they transform under parity transformations [$(\theta,\phi)$ $\rightarrow$ $(\pi-\theta, \phi+\pi)$] between even-parity harmonics, ${\cal E}^{\ell m}$, which transform as: ${\cal E}^{\ell m} \rightarrow (-1)^{\ell}{\cal E}^{\ell m}$, and odd-parity harmonics, ${\cal O}^{\ell m}$, which transform as ${\cal O}^{\ell m} \rightarrow (-1)^{\ell+1}{\cal O}^{\ell m}$.}. Then, for each mode we can decouple the Einstein perturbative equations in terms of \emph{master functions} (combinations of the metric perturbations and their first-order derivatives), which satisfy master wave equations of the form~\footnote{For simplicity we drop the harmonic indices $(\ell,m)$.}:
\begin{equation}
\left(-\frac{\partial^2}{\partial t^2} + \frac{\partial^2}{\partial x^2} - V^{\rm even/odd}_\ell  \right)\Psi^{\rm even/odd} = 0\,,
\label{master-wave-equation}
\end{equation}
where $x$ is the tortoise coordinate, defined by $dx/dr = 1/f$, $V^{\rm even/odd}_\ell(r)$ is the BH potential barrier, and $\Psi^{\rm even/odd}(t,r)$ are the even/odd master functions, which are gauge-invariant (do not depend on the choice of perturbative gauge).
For odd-parity (axial) perturbations, the standard potential considered is the Regge-Wheeler potential~\cite{Regge:1957td} (see also~\cite{Lenzi:2021wpc,Lenzi:2024tgk}):
\begin{equation}
V^{\rm odd}_\ell = V_{\ell}^{\rm RW}(r) = f(r) \left(\frac{\ell(\ell+1)}{r^{2}} - \frac{3r_s}{r^{3}} \right)\,,
\label{RW-potential}
\end{equation}
while for even-parity (polar) perturbations, it is the Zerilli potential~\cite{Zerilli:1970la}:
\begin{eqnarray}
V^{\rm even}_\ell & = & V^{\rm Z}_\ell(r) \nonumber \\[2mm]
& = & \frac{f(r)}{\lambda^{2}(r)} \biggl[ \frac{(\ell-1)^{2}(\ell+2)^{2}}{r^{2}}\left( \ell(\ell+1) + \frac{3r_s}{r} \right) \nonumber \\[2mm]
& + & \frac{9r_s^{2}}{r^{4}}\left( (\ell-1)(\ell+2) + \frac{r_s}{r} \right) \biggr]\,,
\label{Z-potential}
\end{eqnarray}
with $\lambda(r) = (\ell-1)(\ell+2) + 3r_s/r$.
	
Recent work~\cite{Lenzi:2021wpc} has shown that the space of possible master functions and equations is much larger than previously known. Indeed, by assuming that the master functions are linear combinations of the metric perturbations and its first-order derivatives, in~\cite{Lenzi:2021wpc} was found that the space of master functions/equations [of pairs $(V_\ell,\Psi)$] contains two branches: the standard branch, containing the Regge-Wheeler and Zerilli potentials, and the Darboux branch, containing an infinite family of new potentials. All these master equations are connected by DTs~\cite{Lenzi:2021njy}, which are isospectral transformations preserving the reflection and transmission coefficients.  It is also possible to reconstruct the metric perturbations from the master functions in all cases~\cite{Lenzi:2024tgk}.
	
The DT connects elements of the space of master functions and equations (the master equations are determined by the shape of the potential). Then, we say that two pairs $(v,\Phi)$ and $(V,\Psi)$ are connected by a DT, when they are related by the following expressions:
\begin{equation}
\Psi = \Phi_{,x} + g\Phi\,,\qquad V = v + 2g_{,x}\,,
\label{DT-transformations}
\end{equation}
where the DT generating function, $g(x)$, has to satisfy a consistency  in the form of a Riccati equation
\begin{equation}
g^{}_{,x} - g^2 + v = \mathcal{C} \,,
\end{equation}
with $\mathcal{C}$ an arbitrary constant. The DT preserves the frequency spectrum, as it can be seen by working in the frequency domain by introducing the separation $\Psi(t,r) = e^{ikt}\psi(x;k)$, being $k$ the frequency, which leads to a time-independent Schr\"odinger equation
\begin{equation}
\psi^{}_{,xx} - V\psi = -k^2\psi\,.
\label{schrodinger}
\end{equation}
The physical equivalence of all these possible descriptions of the metric perturbations means that the scattering data, in our case the greybody factors as there are no bound states, and the QNM spectrum are the same regardless of which potential is used. This redundancy, what we call \emph{Darboux covariance}~\cite{Lenzi:2021njy}, is at the heart of the developments that follow.

%%%%% KDV SYMMETRIES AND THE BH MOMENT PROBLEM	
\section{KdV Symmetries and the BH Moment Problem}
\label{sec:kdv}
	
The master equations describing BH perturbations possess additional symmetries beyond DTs. These symmetries are generated by the flow of the KdV equation~\cite{Korteweg:10.1080/14786449508620739}
\begin{equation}
V^{}_{,\tau} - 6 V V^{}_{,x} + V^{}_{,xxx} = 0\,,
\label{kdv-equation}
\end{equation}
which is one of the best well-known examples of a completely integrable equation (see~\cite{Ablowitz:1981jq,Zakharov:1991xd,Ablowitz:1991solitons,2002nlin......9057S,Babelon:2003qtg} for diverse literature on integrable systems), a non-linear partial differential equation in this case. Associated to the KdV equation, there is an associated infinite hierarchy of KdV equations that, considered as deformations of our time-independent master equation [Eq.~\eqref{schrodinger}], turn out to be isospectral deformations: they preserve the transmission coefficient as well as the QNM frequencies~\cite{Lenzi:2021njy}. The parameter $\tau$ represents an auxiliary KdV time parameterizing a flow through the space of potentials, distinct from the physical time $t$ of the gravitational perturbations.

Let us now consider a scattering process (as in~\cite{Lenzi:2022wjv}; see ~\cite{Faddeev:1959yc, Faddeev:1976xar, Deift:1979dt, Novikov:1984id, Ablowitz:1981jq} for a detailed treatments of scattering processes), described by our master equations, in which an incident wave comes from the BH horizon, interacts with the BH potential barrier, and part is transmitted and part is reflected. 
The solution of the time-independent Schr\"odinger equation~\eqref{schrodinger} for this physical situation can be written in the following form
\begin{eqnarray}
\psi(x,k,\tau) = \left\{ \begin{array}{lc}
	a(k,\tau) e^{i k x} + b(k,\tau)e^{-i k x}  &  \mbox{for~}x\to -\infty\,,  \\
	&        \\
e^{i k x}  &  \mbox{for~}x \to \infty \,,
\end{array} \right.
\label{plane-wave-ab-deformed}
\end{eqnarray}
where we have introduced the dependence on the KdV time $\tau$ to account for the dependence of the different objects on the KdV deformation of the master equation.  The complex coefficients $a(k,\tau)$ and $b(k,\tau)$ are usually known as the Bogoliubov coefficients and they determine completely the transmission/reflection coefficients (and hence the scattering matrix) as follows 
\begin{eqnarray}
t(k,\tau) = \frac{1}{a(k,\tau)} \,, \quad 
r(k,\tau) = \frac{b(k,\tau)}{a(k,\tau)} \,,
\label{reflection-transmission-coefficient-deformed}
\end{eqnarray}
The transmission and reflection probabilities, i.e. the greybody factors, are simply given by the square of the modulus of the corresponding coefficients
\begin{equation}
T(k,\tau) = |t(k,\tau)|^2\,, \quad R(k,\tau)=|r(k,\tau)|^2 \,,
\label{rt-probabilities-deformed}
\end{equation}
which, for real $k$, satisfy 
\begin{eqnarray}
T(k) + R(k) = 1 \,.
\label{T+R}
\end{eqnarray}
One can show that under the deformations induced by the KdV flow, the Bogoliubov coefficients evolve as~\cite{Gardner:1967wc}
\begin{eqnarray}
a(k,\tau) = a(k,0) \,,\quad  b(k,\tau) = b(k,0) e^{8 i k^3 \tau} \,,
\label{ab-kdv-flow}
\end{eqnarray}
which means that we have a conservation law, namely
\begin{equation}
a^{}_{,\tau}(k,\tau)=0\,. \label{a-conservation-law}
\end{equation}
The modulus of $b(k,\tau)$ is also preserved by the KdV flow. Similar evolution laws, can be obtained for the reflection and transmission coefficients.

The integrability of the KdV hierarchy is related to the presence of an infinite sequence of conserved quantities, the KdV integrals~\cite{Miura:1968JMP.....9.1204M,Zakharov:1971faa,Lax:1968fm}, which in our case can be derived from the conservation law of Eq.~\eqref{a-conservation-law}. They characterize the large-frequency asymptotics of the scattering wave function (see~\cite{Faddeev:1959yc, Faddeev:1976xar, Deift:1979dt, Novikov:1984id, Ablowitz:1981jq} for detailed developments). Writing the solution of the Schr\"odinger equation as (including the KdV flow time $\tau$):
\begin{equation}
\psi(x,k,\tau) = \exp\left\{i k x - \int_{x}^{+\infty} \!\!  \Phi(x',k,\tau)\, dx'\right\} \,,   
\label{phase-ansatz}
\end{equation}
so that in the half-plane, $\Im(k) > 0$, we have
\begin{eqnarray}
a(k, \tau) =  \lim^{}_{x\rightarrow -\infty} e^{-ik x}\, \psi (x,k,\tau)\,,
\label{a-phi}
\end{eqnarray}
and in consequence,
\begin{eqnarray}
\ln{a(k, \tau) }  = -\int_{-\infty}^{\infty}\!\!\!  \Phi(x,k,\tau)\,dx \,.
\label{lna-phi}
\end{eqnarray}
The differential phase function $\Phi$ satisfies the following complex Riccati equation
\begin{equation}
\Phi^{}_{,x} + 2ik\Phi + \Phi^2 = V(x)\,.
\label{riccati}
\end{equation}
Expanding $\Phi$ in inverse powers of $2ik$ (see~\cite{Zakharov:1971faa}),
\begin{equation}
\Phi(x,k,\tau) = \sum_{n=1}^\infty \frac{\kappa^{}_n(x,\tau)}{(2ik)^n}\,,
\label{expansion-phase}
\end{equation}
and introducing it in Eq.~\eqref{riccati} leads to a recurrence for the coefficients $\kappa_n(x,\tau)$:
\begin{eqnarray}
\kappa^{}_1 & = & V\,, \\[2mm] 
\kappa^{}_n & = & -\kappa^{}_{n-1,x} - \sum_{m=1}^{n-2}\kappa^{}_m\kappa^{}_{n-1-m}\,,\quad n\ge 2\,. \label{recurrence}    
\end{eqnarray}
The KdV integrals are conserved with respect to the KdV flow by virtue of Eq.~\eqref{a-conservation-law}, that is, they are independent of the KdV time $\tau$.  They are given by: 
\begin{equation}
\mathcal{K}^{}_n = \int_{-\infty}^{\infty} \kappa^{}_n(x)\,dx\,.
\label{kdv-integrals}
\end{equation}
All even-indexed integrals vanish ($\mathcal{K}_{2n}=0$), while the odd ones are the true conserved quantities.
	
The KdV integrals are invariant under DTs~\cite{Lenzi:2021njy}, meaning they take the same value for every potential in the Schwarzschild perturbation theory equivalence class. Or, in other words, for any element of the space of master equations and functions determined in~\cite{Lenzi:2021wpc}:
\begin{equation}
\mathcal{K}^{}_{2n+1}[V] = \mathcal{K}^{}_{2n+1}[V^{\rm RW}_\ell] = \mathcal{K}^{}_{2n+1}[V^{\rm Z}_\ell]\,,~\forall\,n\ge 0\,.
\label{kdv-invariance}
\end{equation}
Using the analytic properties of $a(k)$ in the upper half $k$-plane ($\Im(k)\geq 0$), one can obtain the trace identities [from the \emph{diagonal} of the resolvent of the master equation~\eqref{schrodinger}]~\cite{Zakharov:1971faa,Lenzi:2022wjv}:
\begin{equation}
\frac{(-1)^{n+1}\mathcal{K}^{}_{2n+1}}{2^{2n+1}} = \frac{1}{2\pi}\int_{-\infty}^{\infty} k^{2n}\ln T(k)\,dk\,.
\label{trace-identities}
\end{equation}
It is straightforward to see that this infinite set of equations define what in mathematics is known as a \emph{classical moment problem}~\cite{Hamburger1920ae,Hausdorff1921,Hamburger1921be,Hamburger1921ce,Riesz1922Part1,Riesz1922Part2,Riesz1923Part3} (see~\cite{akhiezer1965classical,KJELDSEN199319,schmudgen2017moment} for modern accounts). Introducing the positive definite distribution function
\begin{equation}
p(k) = -\frac{\ln T(k)}{2\pi} \ge 0\,,
\label{distribution}
\end{equation}
and the moments
\begin{equation}
\mu^{}_{2j} = (-1)^j\frac{\mathcal{K}^{}_{2j+1}}{2^{2j+1}}\,,
\label{moments-and-KdV}
\end{equation}
the trace identities take a more compact form which clearly corresponds to a moment problem:
\begin{equation}
\mu^{}_{2j} = \int_{-\infty}^{\infty} k^{2j} p(k)\,dk\,,\qquad j=0,1,2,\ldots\,.
\label{moment-problem}
\end{equation}
In this form, this corresponds to a (symmetric~\footnote{In the sense that only the even moments are different from zero.}) Hamburger moment problem~\footnote{Broadly speaking, we can distinguish three types of moment problems: (i) The Hamburger moment problem, where integration is over the real line. (ii) The Stieltjes moment problem, where integration is over the positive real line. (iii) The Hausdorff moment problem, where integration is over a compact interval.}. The goal is to provide the sequence of even moments $\{\mu_{2j}\}_{j=0}^{\infty}$ and, from Eq.~\eqref{moment-problem}, to recover the distribution $p(k)$, from which we obtain the greybody factor $T(k)$. We have numerically checked~\cite{Lenzi:2023inn} that the BH moment problem [as defined by Eq.~\eqref{moment-problem} with the definitions of Eqs.~\eqref{distribution} and~\eqref{moments-and-KdV}] is determinate, meaning a unique solution exists~\cite{schmudgen2017moment,Lenzi:2022wjv} (see also~\cite{baker_graves-morris_1996}).
	
A final remark worth mentioning here concerns the complete integrability of the KdV equation. It was shown in~\cite{Zakharov:1971faa} that there is an infinite-dimensional Hamiltonian system associated (see, e.g.~\cite{Arnold:1989am}) for which a set of action-angle variables, say $(P(k),Q(k))$, exist.  In our case, the action variable $P(k)$ is associated to the distribution $p(k)$ of the BH moment problem [Eq.~\eqref{distribution}] while the angle variable $Q(k)$ is given by the argument of the Bogoliubov coefficient $b(k)$, that is: $P(k) = 2k\, p(k) \,$, and $Q(k) = {\rm arg}\, b(k) \,$.

%%%%% SOLVING THE BH MOMENT PROBLEM	
\section{Solving the BH moment problem}
\label{sec:solving-BH-moment-problem}

In~\cite{Lenzi:2022wjv} there is a broad discussion of basic mathematical aspects of the BH moment problem.  Here, we summarize the main points. First of all, there is a one-to-one relation between the moment problem and the problem of finding the analytic function $F(z)$ in $\mathbb{C\setminus R}$ defined as the Stieltjes transform of the distribution $p(k)$:
\begin{eqnarray}
F(z) = \int_{-\infty}^{\infty} \frac{p(k)}{k-z}\, dk\,,
\label{stieltjes-transform}
\end{eqnarray}
Then, $F(z)$ has the following asymptotic expansion in terms of the moments
\begin{equation}
F(z) = - \sum_{n=0}^{\infty} \frac{\mu^{}_n}{z^{n+1}}  \,,
\label{stieltjes-transform-asymptotic}
\end{equation}
for any range of angles $\delta \leq \arg(z)\leq \pi - \delta$ with $0 < \delta < \pi/2$, can be seen as a generating function for the moments. 
This shows that the distribution $p(k)$ is indeed uniquely determined by $F(z)$, by using the {\em Stieltjes-Perron inversion}~\cite{AFST_1894_1_8_4_J1_0} (see also~\cite{Shohat1943ThePO,akhiezer1965classical,schmudgen2017moment}), which relates the probability distribution with the branch cut discontinuity of the Stieltjes transform across the real axis: 
\begin{eqnarray}
p(k) = \lim^{}_{\epsilon\to 0} \frac{\Im F(k+i\epsilon)}{\pi} \,,
\label{stieltjes-perron-inversion}
\end{eqnarray}
The functions defined by the Stieltjes transform [Eq.~\eqref{stieltjes-transform}] have the Herglotz property~\cite{Bender:1978bo}, $\overline{F(z)} = F(\overline{z})$.

In our case, one can show that~\cite{Lenzi:2022wjv} 
\begin{eqnarray}
\ln{a(k, \tau)} = \frac{1}{2\pi i}\int^{\infty}_{-\infty}\!  \frac{\ln T(k',\tau) }{k-k'}\, dk'\,,
\label{loga-logr}
\end{eqnarray}
which defines a Stieltjes transform and the Stieltjes-Perron inversion becomes 
\begin{eqnarray}
\ln T(k) = - \lim^{}_{\epsilon\rightarrow 0} 2\,\Re \ln{a(k+i\epsilon)} \,.
\label{stieltjes-inversion-loga}
\end{eqnarray}
The practical implementation of this formula is not straightforward as it has been discussed in~\cite{Lenzi:2022wjv}. 
Then, we need to resort to approximate computational methods, which in practice means that we are going to use a finite number of moments in order to compute the transmission coefficient. Solving the moment problem from a finite set of moments is an ill-posed inverse problem: Infinitely many distributions are compatible with a given finite moment sequence.  Also, it is not easy to find robust computational methods to carry out the inversion. 
In Ref.~\cite{Lenzi:2023inn}, we exploited the connection between the moments and the moment generating function. For the Stieltjes formulation of the moment problem, the moment generating function is defined as the Laplace transform of the distribution $\tilde p(\xi)$,
\begin{equation}
M(t)=\int_0^\infty e^{-t\xi}\,\tilde p(\xi)\,d\xi\,,    
\label{definiton-of-MGF}
\end{equation}
where we have introduced the following definitions:
\begin{equation}
\xi = \sigma^2 k^2 \,,  \qquad
\tilde{p}(\xi) = \frac{p(\xi)}{\sqrt{\xi}} = \frac{p(k)}{\sigma\,k} \,,
\label{stieltjes-to-hamburger}
\end{equation}
where $\xi$ is a dimensionless variable and $\sigma=\alpha^{-1}$ for the Pöschl-Teller potential and $\sigma=r_s$ for the Regge-Wheeler one. The asymptotic expansion of the moment generating function for large $t$ is directly given by the moments:
\begin{equation}
M(t)=\sum_{n=0}^{\infty}\frac{m^{}_n}{n!}(-t)^n\,,
\qquad
m^{}_n=\sigma^{2n+1}\mu^{}_{2n}\,.
\end{equation}
Considering a finite set of moments, Pad\'e approximants for the truncated asymptotic series of $M(t)$ were constructed in Ref.~\cite{Lenzi:2023inn}. For Stieltjes series, diagonal and subdiagonal Pad\'e approximants are known to converge to the Stieltjes function [Eq.~\eqref{loga-logr}], which is only valid in the upper half $k$-plane.
Pad\'e approximants of the Stieltjes series of Eq.~\eqref{stieltjes-transform-asymptotic} reproduce the real line branch-cut evident from Eq.~\eqref{loga-logr}.
Given the close connection between both the Stieltjes and the moment generating functions with the distribution $p(k)$, we would also expect that Pad\'e approximants to the second converge, as it happens in the approximations shown in Ref.~\cite{Lenzi:2023inn} and in others quoted there.
The next important step in Ref.~\cite{Lenzi:2023inn} is to decompose the Pad\'e approximant into partial fractions, for which the inverse Laplace transform can be computed. The relevant expressions for the approximate computation of $T(k)$ can be seen in Ref.~\cite{Lenzi:2023inn}

The Pad\'e-based approximation has the advantage of being computationally very efficient and often yields accurate results with only a few moments. However, it may produce poles with positive real parts that must be discarded, and it does not automatically guarantee positivity of the reconstructed distribution at every truncation order (the reconstructed $T(k)$ can locally exceed unity if the exponential argument becomes positive). In practice, these issues are mild for the BH potentials as it was shown in~\cite{Lenzi:2023inn}, and the method compares well with WKB results. 

Before moving to the description of a new method to solve the truncated moment problem, it is important to discuss several general questions that are going to be useful for what comes in the next section. First of all, the physical dimensions (units) of the different quantities involved, taking into account that we are using natural geometric units and hence, all the quantities can be expressed as a power of the length unit, say $L$.  Let us start from the BH potential, which is the object that defines the physical framework. It is clear from the definitions [see Eqs.~\eqref{RW-potential} and~\eqref{Z-potential}] that it has inverse square length dimensions, i.e.
\begin{equation}
[V^{}_{\mathrm{BH}}] = L^{-2} \,.   
\end{equation}
Then, we can see that the dimensions of the KdV densities described in Eq.~\eqref{recurrence} are:
\begin{equation}
[\kappa^{}_{2j+1}] = L^{-2j-2} \,,   
\end{equation}
which means that the dimensions of the KdV integrals in Eq.~\eqref{kdv-integrals} are:
\begin{equation}
[\mathcal{K}^{}_{2j+1}] = L^{-2j-1} \quad \Rightarrow \quad 
[\mu^{}_{2j}] = L^{-2j-1} \,,   
\end{equation}
where we have use the definition of the moments $\mu_{2j}$ in Eq.~\eqref{moments-and-KdV}.
Finally, the frequency, $k$, and the density, $p(k)$, have obvious dimensions:
\begin{equation}
[k] = L^{-1}\,,\qquad
[p(k)] =    1\,.   
\end{equation}
We can introduce dimensionless quantities by using $\sigma$, which was introduced in Eq.~\eqref{stieltjes-to-hamburger}.   
In this way, we can introduce dimensionless moments and frequency as follows:
\begin{equation}
\tilde{\mu}^{}_{2j} = \mu^{}_{2j} \sigma^{2j+1}\quad
(j=0,1,\ldots)\,,\quad  \tilde{k} = \sigma k \,.
\label{transformation-dimensionless}
\end{equation}
%

%%%%% A NEW METHOD: MAXIMUM ENTROPY RECONSTRUCTION
\section{A New Method: Maximum Entropy Reconstruction}
\label{sec:maxent}

A plausible alternative to the Pad\'e-based method described above is the maximum entropy (MaxEnt) method~\cite{Jaynes1957a,Meadpapanicolau1984}.  Mathematical aspects of the method, like existence, convergence and numerical implementations have been discussed, for instance in~\cite{BorweinLewis1991,TAGLIANI1999291,Abramov2010,BiswasElliott2026}.  The MaxEnt method has been used previously for the moment problem in other physical contexts, among others: Electronic densities of states~\cite{Turek1988,DraboldSankey1993,SilverRoder1997,Bandyopadhyay2005}; reconstruction of parton and baryon-number distributions from moments and cumulants~\cite{Zhang2024}; polymer distribution functions~\cite{Poland1995}; invariant densities of chaotic maps~\cite{DingMead2002,BiswasShimoyamaMead2010}; entropy-based moment closures in kinetic theory~\cite{Levermore1996,HauckLevermoreTits2008}; etc.

The MaxEnt method~\cite{Meadpapanicolau1984} selects the least-biased distribution consistent with the
known moments by maximizing the Shannon entropy
\begin{equation}
S[p(x)] = -\int p(x)\ln p(x)\,dx\,.
\label{entropy}
\end{equation}
Consider the classical moment problem for a positive density $p(x)$ on $[a,b]$,
\begin{equation}
\mu^{}_{n} = \int_a^b x^{n} p(x)\,dx\,,\qquad n=0,1,\ldots,N\,.
\end{equation}
The MaxEnt method maximizes the functional
\begin{eqnarray}
S[p(x)] & = & -\int_a^b\left[p(x)\ln p(x)-p(x)\right]dx \nonumber \\[2mm]
  & -  & \sum_{n=0}^N\lambda_{n}\left(\int_a^b x^{n}p(x)\,dx - \mu_{n}\right),
\end{eqnarray}
where the $\lambda_n$ are Lagrange multipliers. Since
only a finite number of moments is known, we can only reconstruct an approximation $p_N(x)$ to $p(x)$.
The functional variation with respect to $p$ yields
\begin{equation}
p^{}_N(x)
=\exp\left\{ -\lambda^{}_0 - \sum_{n=1}^N \lambda^{}_{n} x^{n}\right\}\,.
\end{equation}
To write $\lambda_0$ in terms of the remaining multipliers, we have to normalize all the moments by
$\mu_0$, so that the zeroth moment equals one.

Let us now turn to our moment problem. Since the distribution $p(k)$ in Eq.~\eqref{moment-problem} can be
taken to be even, we fold the problem onto the half-line by introducing
\begin{equation}
    q(k)=2p(k) \,,
\label{q-distribution}
\end{equation}
and hence, we have the following Stieltjes moment problem:
\begin{equation}
\mu_{2j}=\int_0^\infty k^{2j} q(k)\,dk\,,\qquad j=0,1,\ldots,N\,.
\label{half-line-moments}
\end{equation}
Normalizing by
$\mu_0$, the moment problem becomes
\begin{equation}
\hat{\mu}^{}_{2j}=\frac{\mu^{}_{2j}}{\mu^{}_0} = \int_0^\infty k^{2j} Q(k)\,dk\,,\qquad j=0,1,\ldots,N\,,
\label{norm-moments}
\end{equation}
where $[\hat{\mu}_{2j}]=L^{-2j}$ and $Q(k)=q(k)/\mu_0$, so that $\hat{\mu}_0=1$. These equations are the constraints for the maximization
of the entropy, introduced through Lagrange multipliers $\lambda_{2j}$, one for each of the $N+1$ retained
moments. The functional to be extremized is
\begin{eqnarray}
S[Q(k)] & = & -\int_0^\infty\left[Q(k)\ln Q(k)-Q(k)\right]dk \nonumber \\[2mm]
  & +  & \sum_{j=0}^N\lambda_{2j}\left(\int_0^\infty k^{2j}Q(k)\,dk - \hat{\mu}_{2j}\right)\,,
\end{eqnarray}
and its stationary point is
\begin{equation}
Q^{}_N(k)=\exp\left\{ -\lambda^{}_0 - \sum_{j=1}^N \lambda^{}_{2j} k^{2j}\right\}\,.
\label{QN}
\end{equation}
Since $\hat{\mu}_0=1$, $Q_N$ must be normalized to one, which fixes $\lambda_0$ in terms of the other multipliers:
\begin{equation}
\lambda^{}_0 = \ln Z\,,\qquad
Z = \int_0^\infty \exp\left\{ -\sum_{j=1}^N \lambda^{}_{2j} k^{2j}\right\}\, dk\,.
\label{partition-function}
\end{equation}
The multipliers $\lambda_{2j}$ ($j=1,\ldots,N$) are found by minimizing the dual potential
\begin{equation}
\Gamma(\lambda_2,\ldots,\lambda_{2N}) = \ln Z + \sum_{j=1}^N \hat{\mu}_{2j}\lambda_{2j}\,.
\label{dual-potential}
\end{equation}
$\Gamma$ is convex in the multipliers, and its gradient is
\begin{equation}
\frac{\partial\Gamma}{\partial\lambda^{}_{2j}} = \hat{\mu}^{}_{2j} - \langle k^{2j}\rangle\,,
\label{gradient}
\end{equation}
where the average is taken with $Q_N$, so its stationary points coincide with the solutions of the moment constraints. We minimize $\Gamma$ using the L-BFGS-B algorithm~\cite{Byrd1995,Zhu1997}, a limited-memory quasi-Newton method with constraints.

The highest-order multiplier $\lambda_{2N}$ is constrained to be strictly positive, which guarantees the convergence of the integral defining $Z$. Since this integral extends to infinity, in practice we truncate it at a finite value of $k$, chosen large enough for the integrand to be negligible beyond it.
In addition, we clip the exponent inside the integrand to avoid numerical overflow.\footnote{By this we mean that, in order to prevent numerical overflow, we limit the argument of the exponential in the integrand from Eq.~\eqref{partition-function} to $[-c, c]$. While double-precision $e^x$ overflows at $x \gtrsim 709.78$ (and underflows safely to zero at $\lesssim -745.13$), quadrature accumulation causes the integral to overflow earlier, so $c$ is chosen with a safety margin. This numerical safeguard does not alter the converged result.}

Before normalizing and minimizing, we rescale the moments, since they differ from each other by orders of magnitude.
We introduce the scale
\begin{equation}
k^{}_s = \hat{\mu}_{2N}^{\frac{1}{2N}}=\left(\frac{\mu^{}_{2N}}{\mu^{}_0}\right)^{\frac{1}{2N}}\,,
\qquad [k^{}_s] = L^{-1}\,,
\label{k_s}
\end{equation}
that is, it has the units of $k$, so that
\begin{equation}
\nu=\frac{k}{k^{}_s}\,,
\label{nu-definition}
\end{equation}
is a dimensionless variable. The rescaled moments and multipliers are
\begin{equation}
\mu'^{}_{2j} = \frac{\mu^{}_{2j}}{k_s^{2j}}\,,\quad
\lambda'^{}_{2j} = k_s^{2j}\lambda^{}_{2j}\,,\quad \lambda'_0=\lambda_0-\ln\,k^{}_s\,.
\label{rescaling}
\end{equation}
By construction, $\hat{\mu}'_0=\hat{\mu}'_{2N}=1$, and by the log-convexity of the moments all the others
lie in $(0,1]$. 
The normalized constraints are: $\hat{\mu}'_{2j}=\langle \nu^{2j}\rangle$, which enter
in the expression of $\Gamma$. Since $\mu'_0=\mu_0$, rescaling and normalizing commute. The minimization of $\Gamma$ is carried out entirely with respect to the variable $\nu$.

Once the rescaled multipliers $\lambda'_{2j}$ are found, $\lambda'_0$ follows from the normalization
condition:
\begin{equation}
\lambda'^{}_0 = \ln\left[\int_0^\infty \exp\left\{-\sum_{j=1}^N \lambda'^{}_{2j}\nu^{2j}\right\}\,d\nu\right]\,,
\label{eq:l_0}
\end{equation}
so that the rescaled density, normalized to one in the variable $\nu$, is
\begin{equation}
Q'^{}_N(\nu)=\exp\left\{-\lambda'_0 - \sum_{j=1}^N \lambda'_{2j}\,\nu^{2j}\right\}\,.
\end{equation}
The rescaling is undone only when reconstructing the greybody factor. Since $k=k_s\nu$,
the density in the original variable is obtained by evaluating $Q'_N$ at $\nu=k/k_s$ and
dividing by $k_s$ (the Jacobian of the change of variable):
\begin{equation}
Q^{}_N(k)=\frac{1}{k_s}\,Q'_N\!\left(\frac{k}{k_s}\right)\,.
\end{equation}
Finally, we have that the greybody factor from Eq.~\eqref{distribution} is
\begin{equation}
\begin{split} 
T^{}_N(k) &=\exp\left\{-\pi\,\mu_0\,Q^{}_N(k)\right\}\\
        &= \exp\left[-\frac{\pi\,\mu^{}_0}{k^{}_s}\exp\left\{-\lambda'^{}_0
- \sum_{j=1}^N \lambda'^{}_{2j}\left(\frac{k}{k^{}_s}\right)^{2j}\right\}\right]\,.
\end{split}
\label{reconstructed-T}
\end{equation}
Equivalently, one can obtain $Q_N(k)$, Eq.~\eqref{QN}, in terms of the original multipliers, using Eq.~\eqref{rescaling}.

The MaxEnt method has several advantages to deal with the moment problem: 
(i) it guarantees positivity of the reconstructed distribution at every order of truncation; 
(ii) the reconstruction is an entire function of $k$, with no spurious poles; 
(iii) the method provides a systematic way to incorporate additional moments as they become available.

%%%%% VALIDATION OF THE MAXENT METHOD: THE POSCHL-TELLER POTENTIAL
\section{Validation of the MaxEnt Method: The P\"oschl-Teller Potential}
\label{sec:pt}
	
Before applying the MaxEnt method to the BH potential barriers, we are going to validate it against the P\"oschl-Teller potential~\cite{Poschl:1933zz}, one of the few exactly solvable potentials on the real line:
\begin{equation}
V^{}_{\rm PT}(x) = \frac{V^{}_0}{\cosh^2\left[\alpha(x-x^{}_0)\right]} 
                 = \frac{\alpha^2(\beta^2 + 1/4)}{\cosh^2\left[\alpha(x-x^{}_0)\right]}\,,
\label{PT-potential}
\end{equation}
where $\beta = \sqrt{V_0/\alpha^2 - 1/4} > 0$.  
It is important to mention that the P\"oschl-Teller potential has been frequently used as an approximation to the actual BH potential. The main reason is that the P\"oschl-Teller potential, as it has been written in Eq.~\eqref{PT-potential} with $V_0>0$, is also a positive potential barrier. While the BH and P\"oschl-Teller potentials decay exponentially as we approach the horizon ($x\rightarrow -\infty$), the main difference between them lies in the behaviour at spatial infinity ($x\rightarrow\infty$), where the P\"oschl-Teller potential also decays exponentially while the BH potentials decay as a power law, specifically: $V_\mathrm{BH} \sim \ell(\ell+1)/r^{2}$. That is, the behavior at spatial infinity is dominated by the centrifugal barrier.

Taking into account the similar qualitative behaviour between the P\"oschl-Teller potential and the BH potential, the first one has been used in the literature~\cite{Ferrari:1984ozr,Ferrari:1984zz} as an approximation to the BH potential barrier. To that end, the first step is to identify the coordinate $x$ in Eq.~\eqref{PT-potential} with the tortoise coordinate of the Schwarzschild BH. Then, the value of the constant $x_0$ is chosen to be the location of the maximum of the BH potential and hence, $V_0$ is identified with the maximum value of the BH potential, that is, $V'_\mathrm{PT}(x_0)=0$ and $V_\mathrm{PT} = V_0 = \max(V_\mathrm{BH})$. Finally, the constant $\alpha$ in Eq.~\eqref{PT-potential} is related to the BH potential curvature via: $\alpha^2 = - V''_\mathrm{BH}(x_0)/(2V_0)$.

The exact transmission coefficient for the P\"oschl-Teller potential can be found analytically. The expression is:
\begin{equation}
T^{}_{\rm PT}(k) = \frac{\sinh^2\left(\frac{\pi k}{\alpha}\right)}{\cosh^2\left(\frac{\pi k}{\alpha}\right) + \sinh^2\left(\pi\beta\right)}\,.
\label{PT-exact}
\end{equation}

Since $T^{}_{\rm PT}(-k)=T^{}_{\rm PT}(k)$, the associated distribution is also symmetric, $p(-k)=p(k)$, and we can restrict ourselves to positive frequencies, as in Eq.~\eqref{half-line-moments}. For the P\"oschl-Teller potential the natural length scale is $\sigma=\alpha^{-1}$, so the dimensionless variables of Eq.~\eqref{transformation-dimensionless} are
\begin{equation}
\tilde{k}=\frac{k}{\alpha}\,,\quad
\tilde{V}^{}_{\rm PT} = \frac{V^{}_{\rm PT}}{\alpha^2}\,,
\label{PT-dimensionless}
\end{equation}
which satisfy $k\,x=\tilde{k}\,\tilde{x}$. For convenience, we have also introduced the following definitions:
\begin{equation}
\tilde{x}=\alpha\,x\,,\quad
\tilde{x}^{}_0=\alpha\,x^{}_0\,.
\end{equation}
Then, the Schr\"odinger equation, Eq.~\eqref{schrodinger}, becomes $\psi_{,\tilde{x}\tilde{x}}-\tilde{V}\psi=-\tilde{k}^2\psi$, and the potential and the greybody factor take the form
\begin{eqnarray}
\tilde{V}^{}_{\rm PT}(\tilde{x}) & = & \frac{\beta^2 + 1/4}{\cosh^2(\tilde{x}-\tilde{x}^{}_0)}\,,
\label{V_pt_tilde}\\[2mm]
T^{}_{\rm PT}(\tilde{k}) & = & \frac{\sinh^2(\pi\tilde{k})}{\cosh^2(\pi\tilde{k}) + \sinh^2(\pi\beta)}\,,
\label{T_pt_real}
\end{eqnarray}
where we have used $k/\alpha = \tilde{k}$ in Eq.~\eqref{PT-exact}. The shift $\tilde{x}_0$
does not affect the scattering data, since the potential only depends on $\tilde{x}-\tilde{x}^{}_0$.  

Finally, the dimensionless moments are the ones from Eq.~\eqref{transformation-dimensionless}.
Multiplying Eq.~\eqref{half-line-moments} by $\sigma^{2j+1}$ and using that the density $q$ is dimensionless,
the moment problem becomes
\begin{equation}
\tilde\mu^{}_{2j} = \int^\infty_{0} \!\!\tilde{k}^{2j}\,q(\tilde{k})\,d\tilde{k}\,,
\label{mu_2j_tilde}
\end{equation}

The first KdV integrals for the P\"oschl-Teller potential are given in Appendix~\ref{app:kdv-pt}.  Using these exact values for the moments we have carried out the inversion of the moment problem as described in Sec.~\ref{sec:maxent}.

%%%%% FIGURE
\begin{figure}[ht!]
\centering
\includegraphics[width=0.95\linewidth]{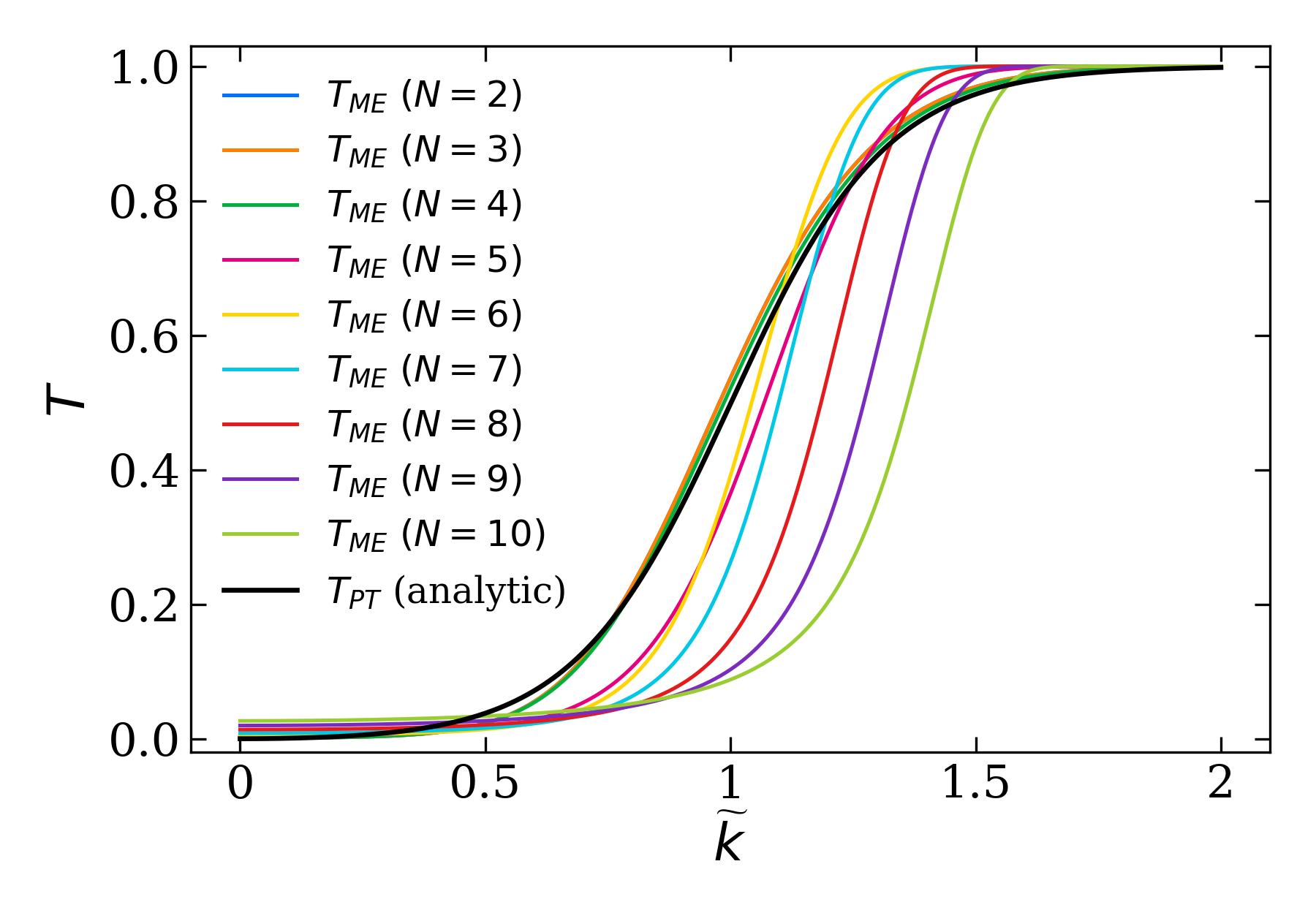}
\vspace{0.2cm}
\includegraphics[width=0.95\linewidth]{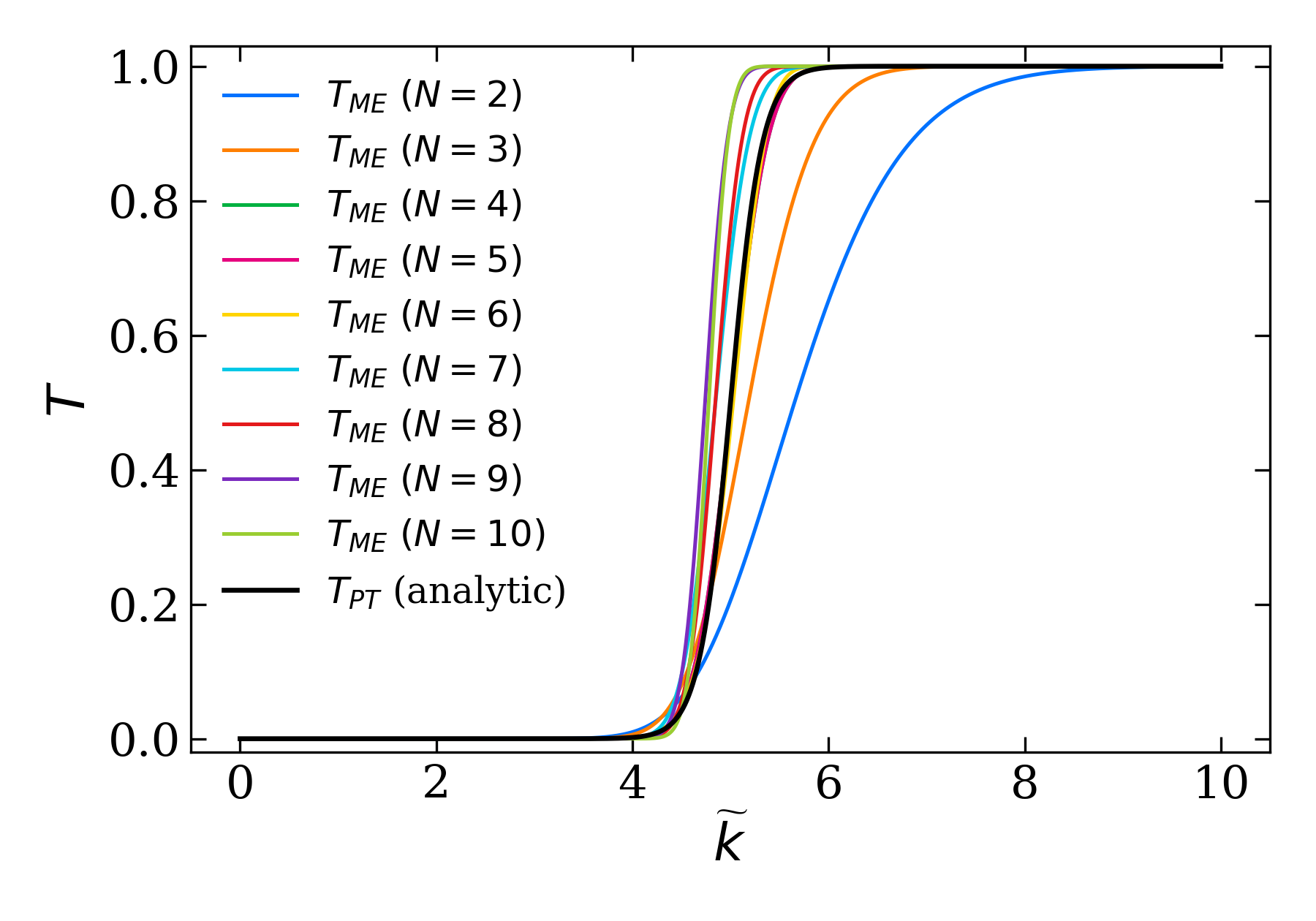}
\vspace{0.2cm}
\includegraphics[width=0.95\linewidth]{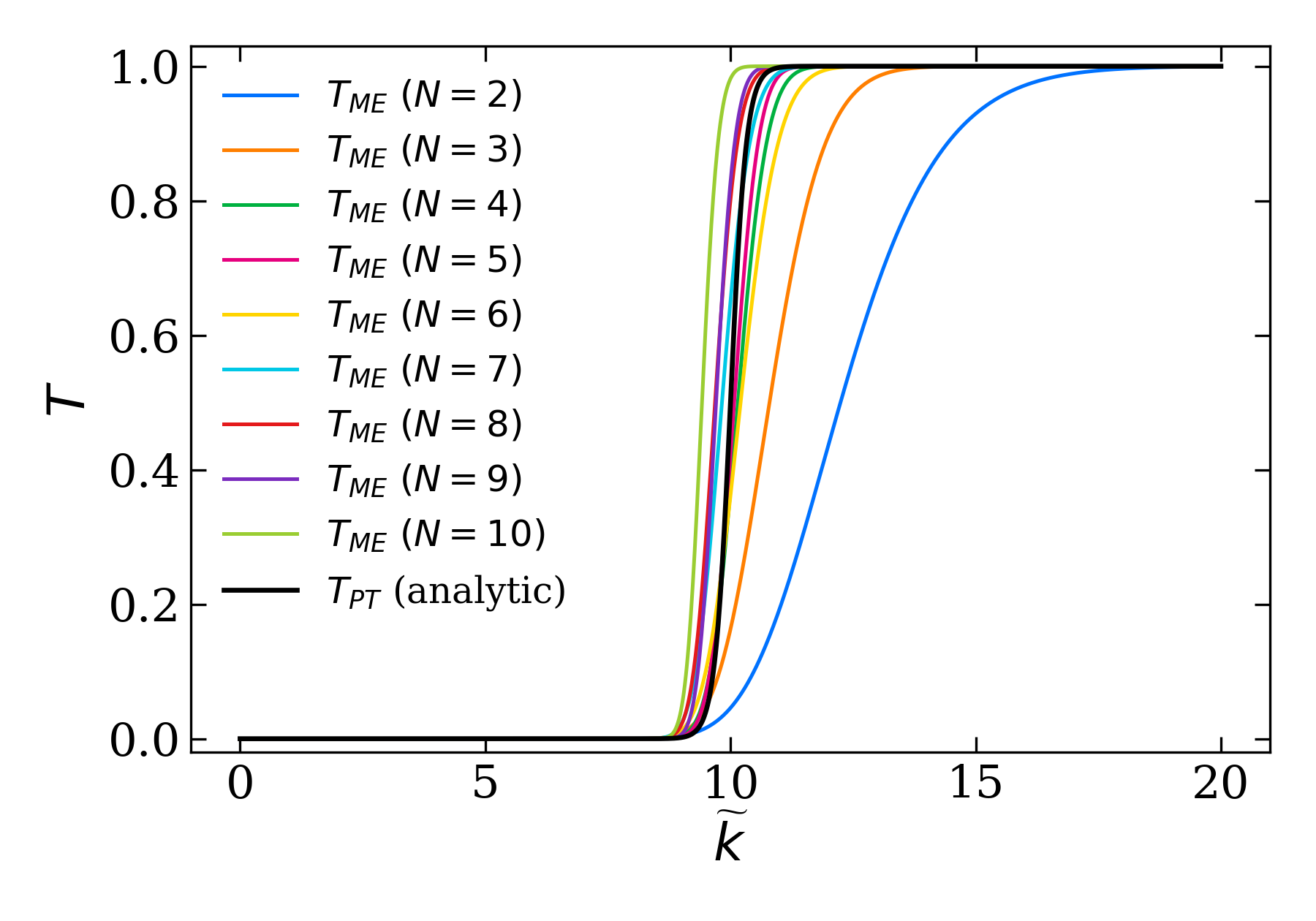}
\caption{Greybody factors for the P\"oschl-Teller potential as functions of the dimensionless 
%wave number 
frequency $\tilde{k}$, obtained from the MaxEnt reconstruction described in Sec.~\ref{sec:maxent}, and compared with the exact solution [Eq.~\eqref{PT-exact}], for $\beta=1$ (top), $\beta=5$ (middle), and $\beta=10$ (bottom). The subscript $N$ indicates the number of moments used in the reconstruction.}
\label{fig:pt-reconstruction}
\end{figure}
%%%%%

\Cref{fig:pt-reconstruction} shows the MaxEnt reconstruction, $T_N(\tilde{k})$, for different values of the P\"oschl-Teller constant $\beta$ and the number of moments considered, $N$. The reconstruction captures the 
%qualitative 
shape of $T_{\rm PT}(\tilde{k})$ even for small values of $N$. 
However, the number of moments required to obtain a faithful reconstruction depends strongly on $\beta$. For $\beta=1$, the curves with $N=2,3,4$ are already very close to the
exact result (the curves for $N=2$ and $N=3$ are indistinguishable). For $\beta=5$ and $\beta=10$, by contrast, the low-order reconstructions ($N=2,3$) exhibit an overly gradual transition, requiring at least $N=4,5$ moments (even $N=6$ moments for $\beta=5$) to accurately capture the correct steepness, position, and width of the step.
	
A surprising feature of~\Cref{fig:pt-reconstruction} is the fact that increasing the number of moments does not seem to improve the reconstruction. For $\beta=1$, the curves with $N>4$ are clearly worse: they undershoot the exact result around the step and overshoot it beyond. 
For $\beta=5$ and $\beta=10$ the best agreement is reached at intermediate $N$ ($N\simeq4$--$6$ and $N\simeq4$--$5$, respectively). Beyond that, the reconstruction becomes steeper than the exact
transmission coefficient and it is displaced with respect to it, towards
smaller $\tilde k$ for $\beta=5$ and $\beta=10$, with a displacement
that grows with $N$ (for $\beta=1$ the displacement of the high-$N$
curves is in the opposite direction, towards larger $\tilde k$). Hence, the best reconstruction is obtained at an intermediate number of moments, and adding further moments degrades rather than refines the result.
	
The results above rely on several truncations that have no counterpart in the exact problem: the upper limit in the integral defining $Z$, the clipping of the exponent, and the bounds imposed on the multipliers during the minimization. It is therefore difficult to separate the limitations of the MaxEnt method itself from those introduced by our numerical implementation. In the next section, we reformulate the problem on a compact interval, which removes the need for the truncations of integrals and allows us to approach the exact result as closely as the method permits.

%%%%% IMPROVING THE BH MAXENT METHOD: COMPACTIFICATION
\section{Improving the BH MaxEnt Method: Compactification}	
\label{sec:compact}

One of the main difficulties involved in inverting the BH moment problem is that the integrals defining the moments have an infinite range of integration, both in the physical and in the frequency domains. In the case of the physical domain, it is not a problem as we have already shown that the KdV integrals for the potentials under consideration can be obtained analytically.
In the case of the frequency domain, we have been dealing either with a Hamburger or with a Stieltjes moment problem. In both cases we have unbounded integration domains.  
Given that we have to truncate the integration domain in the numerical implementations of our algorithm, this can induce errors that will depend on the way we truncate.  
The best we can expect is that the contributions from the part of the domain that has been stripped away do not contribute much to the momenta.  
However, it is also possible that this truncation may affect the convergence of the maximization process of the MaxEnt method, in the sense that the algorithm can at some point be driven to favor terms whose contribution is canceled by the truncation.

In the case of P\"oschl-Teller potential, we can deal better with this problem as it decays exponentially towards the two ends of the domain.  As we have already mentioned, the situation is different for a BH potential, for instance for the Regge-Wheeler potential, as it only decays exponentially towards the horizon, but it decays as a negative power of $r$ (or $x$) towards spatial infinity, where the centrifugal barrier dominates, i.e. $V_\mathrm{RW} \sim \ell(\ell+1)/r^2$.   
In this sense, it would be very convenient to transform the integration domain to a finite interval, that is, to convert our Hamburger/Stieltjes moment problem into a Hausdorff moment problem, or to a similar formulation.  And this is what we explore in this section. 
	
Our starting point is the Hamburger moment problem:
\begin{equation}
\mu^{}_{2j}  =  \int^\infty_{-\infty}\!\! k^{2j}\,p(k)\,dk\,,
\end{equation}
where $j=0,1,\ldots$, the even moments are connected with the BH KdV integrals by Eq.~\eqref{moments-and-KdV}, and the (probability) density function $p(k)$ by Eq.~\eqref{distribution}.
Recall that the KdV integrals, $\mathcal{K}_n$, are conserved quantities of the KdV flow which can be computed from the KdV densities, $\kappa_n$, through Eq.~\eqref{kdv-integrals}. The KdV densities are obtained by means of the recurrence given in Eq.~\eqref{recurrence}. Their analytic expressions for the P\"oschl--Teller and Regge--Wheeler potentials are given in Appendices~\ref{app:kdv-pt} and~\ref{app:kdv-rw}, respectively.

We have seen before how to introduce dimensionless variables and parameters [see Eq.~\eqref{transformation-dimensionless}] and how to restrict the integration to positive frequencies [see Eq.~\eqref{half-line-moments}]. This leads to the formulation of the moment problem in Eq.~\eqref{mu_2j_tilde}, which is very similar to a Stieltjes moment problem. To bring it to that form, we change variable in the dimensionless frequency,
\begin{equation}
\tilde{k}\rightarrow u = \tilde{k}^2 \quad\longleftrightarrow\quad \tilde{k}=\sqrt{u}
\ \Rightarrow\ d\tilde{k} = \frac{d u}{2\sqrt{u}} \,,
\end{equation}
which gives the Stieltjes moment problem
\begin{equation}
m^{}_{j} = \int^\infty_{0} d u\, u^{j}\,g(u)\,, \qquad j=0,1,\ldots\,,
\end{equation}
where [see Eq.~\eqref{q-distribution}]
\begin{equation}
m^{}_j \equiv \tilde{\mu}^{}_{2j}
\quad \mathrm{and} \quad
g(u) = \frac{q(\sqrt{u})}{2\sqrt{u}}\,,
\end{equation}
that is, we have an Stieltjes moment problem with dimensionless moments [see Eq.~\eqref{transformation-dimensionless}].

The key point to transition towards a compactification that produces integrals on a compact interval is
a change of integration variable. We use a relatively simple one,
\begin{equation}
u = \frac{w}{1-w} \quad \longleftrightarrow\quad w = \frac{u}{1+u}=\frac{\tilde{k}^2}{1+\tilde{k}^2}\,,
\end{equation}
which maps $[0,\infty)$ into $[0,1)$. The moments are then given by
\begin{eqnarray}
m^{}_j = \int^1_0 \!\frac{w^j}{(1-w)^{j+2}}\, g\left(\frac{w}{1-w} \right)\, dw \,.
\label{finite-integral-one}
\end{eqnarray}
In practice, to solve numerically the moment problem we can only use a finite number of moments,
parametrized by $N$, i.e.~$j=0,\ldots,N$. Then Eq.~\eqref{finite-integral-one} can be rewritten as
\begin{eqnarray}
m^{}_j = \int^1_0 \! w^j(1-w)^{N-j}\, h^{}_N(w)\,d w \,,
\label{finite-integral-two}
\end{eqnarray}
where
\begin{equation}
h^{}_N(w) = \frac{1}{(1-w)^{N+2}}\, g\left(\frac{w}{1-w} \right) \,.
\end{equation}
The factor in front of $h_N(w)$ is related to the appearance of the Bernstein polynomials in the integrals of our moment problem [Eq.~\eqref{finite-integral-two}]. The Bernstein polynomials are typically defined as
\begin{equation}
B^{}_{N,j}(x) = \binom{N}{j} x^j (1-x)^{N-j} \,.
\label{Bernstein-polynomials}
\end{equation}
Given the connection of the Bernstein polynomials with the binomial formula
\begin{equation}
(a + b)^N = \sum_{j=0}^N \binom{N}{j}a^j b^{N-j} \,,
\end{equation}
and since in our case $a=w$ and $b=1-w$, so that $a+b=1$, we have
\begin{equation}
\sum_{j=0}^N B^{}_{N,j}(w)=1 \,.    
\label{partition-unit-bernstein-polynomials}
\end{equation}

Multiplying Eq.~\eqref{finite-integral-two} by $\binom{N}{j}$ and summing over $j$ gives the identity
\begin{equation}
\sum_{j=0}^N \binom{N}{j}\, m^{}_j = \int^1_0 \! h^{}_N(w)\, dw  \equiv \Sigma^{}_N  \,.
\end{equation}
We normalize the moments and the distribution by $\Sigma_N$ %instead of $\tilde{m}'_0$:
\begin{equation}
M^{(N)}_j =  \binom{N}{j}\frac{m^{}_j}{\Sigma^{}_N}\,,\qquad H^{}_N(w)=\frac{h^{}_N(w)}{\Sigma^{}_N}\,,
\label{normalized_moments_Mj}
\end{equation}
which satisfy
\begin{eqnarray}
\sum_{j=0}^N M^{(N)}_j = 1 \,,\quad
\int^1_0 \! H^{}_N(w)\,dw = 1\,.
\end{eqnarray}
Since the new moments, $\{M^{(N)}_j\}$, are positive we must have
\begin{equation}
   0 \leq  M^{(N)}_j \leq 1 \,. 
\label{M_j01}
\end{equation}
The resulting generalized moment problem is
\begin{equation}
M^{(N)}_j = \int^1_0 \! B^{}_{N,j}(w)\, H^{}_N(w)\, dw\,, \quad j=0,\ldots,N\,.
\label{M_jmomentp}
\end{equation}

The entropy functional from the MaxEnt method that needs to be applied to this form of the moment problem is
\begin{eqnarray}
&&S[H^{}_N(w)] = -\int_{0}^{1} \left[ H^{}_N(w) \ln H^{}_N(w) - H^{}_N(w) \right] dw 
\nonumber \\[2mm]
& & + \sum_{j=0}^{N} \lambda^{}_{j} \left( \int_{0}^{1} B^{}_{N,j}(w) H^{}_N(w) \, dw - M^{(N)}_j \right)\,.
\label{entropy_for_HN}
\end{eqnarray}
and the solution is
\begin{equation}
H^{}_N(w) = \exp\left[-\lambda^{}_0 (1-w)^N - \sum_{j=1}^{N} \lambda^{}_{j}\,B^{}_{N,j}(w)
\right]\,.
\label{distribution_HN}
\end{equation}
Notice that now $\lambda_0$ cannot be expressed in terms of the remaining Lagrange multipliers, so we
have to treat all of them at once. Instead of $\Gamma$, we minimize the dual potential
\begin{eqnarray}
& & \Delta = \int_0^1 \biggl( \exp \Bigl[ -\lambda^{}_0 (1-w)^N \nonumber\\[2mm]
& &  - \sum_{j=1}^{N} \lambda^{}_{j}\,B^{}_{N,j}(w)  \Bigr] - 1 \biggr) dw + \sum_{j=0}^{N}\lambda^{}_{j} M^{(N)}_{j}\,.
\label{Delta}
\end{eqnarray}

The moment problem defined by Eq.~\eqref{M_jmomentp}, with moments that have the property in Eq.~\eqref{M_j01}, looks very appealing to be solved by the MaxEnt method [see Eqs.~\eqref{entropy_for_HN} and~\eqref{distribution_HN}]. However, the fact that the normalized moments $\{M^{(N)}_j\}$ are bound to live in the interval $[0,1]$ does not guarantee that the numerical algorithms to obtain the MaxEnt solution are going to converge. Indeed, we have observed in our numerical experiments, that despite the moments belong to the interval $[0,1]$ they can still differ by orders of magnitude. Therefore, as it is, this generalized formulation of the moment problem is not amenable for numerical computations. 

To revert the situation, following what we did in previous sections, we introduce a rescaling of the moments.  The rescaling factor is defined as follows
\begin{equation}
\rho \equiv \left(\frac{M^{(N)}_N}{M^{(N)}_0}\right)^{\frac{1}{2N}}
= \left(\frac{m_N}{m_0}\right)^{\frac{1}{2N}}\,,
\label{eq:rho}
\end{equation}
where we have used Eq.~\eqref{normalized_moments_Mj}. This factor is dimensionless as the moments we are dealing with are also dimensionless.  In this sense, the factor $\rho$ we have just introduced is going to play the same roles as the scale $k_s$ in Sec.~\ref{sec:maxent}, with the difference that $k_s$ was not dimensionless but has the units of frequency. 
In practice, we implement now the rescaling as a simplified
M\"obius transformation of the variable $w$ that we used for the compactification of the frequency domain. That is, we introduce
\begin{equation}
w = \frac{\rho^{2}\,\omega}{1+(\rho^{2}-1)\,\omega}\,,
\label{eq:mobius}
\end{equation}
and the inverse transformation is, as expected, another M\"obius transformation:
\begin{equation}
\omega = \frac{w}{\rho^{2}+(1-\rho^2)w}  = \frac{\tilde{k}^{2}}{\tilde{k}^{2}+\rho^{2}}\,,
  \label{eq:mobius-inverse}
\end{equation}
where $\tilde{k}$ is the dimensionless frequency introduced in Eq.~\eqref{transformation-dimensionless}. Given that $w\in[0,1]$, the transformation of Eq.~\eqref{eq:mobius}, and its inverse of Eq.~\eqref{eq:mobius-inverse}, are mappings of this interval onto itself.  Other relevant properties of the M\"obius map are:
\begin{equation}
dw = \rho^{2}\,\frac{d\omega}{\mathcal{D}^{2}}\,,\qquad 
1-w=\frac{1-\omega}{\mathcal{D}}\,,
\label{dw}
\end{equation}
where
\begin{equation}
\mathcal{D} = 1 + (\rho^2-1)\omega \,.    
\end{equation}
Now, we can introduce rescaled moments as
\begin{equation}
M'^{(N)}_j=\frac{M^{(N)}_j}{\rho^{2j}}\,.
\label{rescaling_compact}
\end{equation}
However, the M\"obius map does not preserve the Bernstein basis. From their definition in Eq.~\eqref{Bernstein-polynomials}, we find that
\begin{equation}
B^{}_{N,j}(w) = \frac{\rho^{2j}}{D^{N}}\, B^{}_{N,j}(\omega)\,,
\label{eq:B_mobius}
\end{equation}
from where, using the property of Eq.~\eqref{partition-unit-bernstein-polynomials}, we have
\begin{equation}
\mathcal{D}^{N} = \sum_{i=0}^{N}\rho^{2i}B^{}_{N,i}(\omega)\,.
\label{eq:B_mobius-2}
\end{equation}
This is the key identity to rewrite the momentum constraints of Eq.~\eqref{M_jmomentp}. Indeed, using Eqs.~\eqref{dw} and~\eqref{eq:B_mobius} we can write
\begin{equation}
M'^{(N)}_j = \int_0^1 B_{N,j}(w')\,
\frac{\rho^{2}}{\mathcal{D}^{N+2}}\,H^{}_N\big(w(\omega)\big) \,d\omega\,,
\label{eq:constraints_rescaled_raw}
\end{equation}
Now, we can introduce the rescaled distribution as follows:
\begin{equation}
H'^{}_N(\omega) = \frac{1}{\Sigma'^{}_N}\frac{\rho^2}{\mathcal{D}^{N+2}}H^{}_N(w(\omega))\,,   
\end{equation}
with
\begin{equation}
\Sigma'^{}_N =   \sum_{j=0}^{N} M'^{(N)}_j
= \rho^{2}\int_0^1 \frac{H^{}_N(w(\omega))}{\mathcal{D}^{N+2}}\,d\omega \,,
\label{eq:Sigma_tilde}   
\end{equation}
so that the new rescaled distribution is normalized:
\begin{equation}
\int_0^1 H'^{}_N(\omega)\,d\omega = 1\,.
\end{equation}
We can introduce another set of moments that add to unity. Indeed,
\begin{equation}
\hat{M}^{(N)}_j = \frac{M'^{(N)}_j}{\Sigma'^{}_N} \,, \qquad 
\sum_{j=0}^{N}\hat{M}^{(N)}_j= 1\,.
\label{eq:Mprime}
\end{equation}
Then, after the rescaling with the factor $\rho$ we have arrived at the same type of generalized moment problem as in Eq.~\eqref{M_jmomentp}:
\begin{equation}
\hat{M}^{(N)}_j = \int_0^1 B_{N,j}(\omega)\,H'^{}_N(\omega)\,d\omega\,, 
\quad j=0,\dots,N \,.
\label{eq:moment_problem_rescaled}
\end{equation}
In addition, by construction, we have
\begin{equation}
\hat{M}^{(N)}_0 = \hat{M}^{(N)}_N \,,
\end{equation}
so that the new moments are no going to be separated by orders of magnitude.

The numerical procedure to invert this generalized moment problem starts with the computation of the dimensionless raw moments $m_j=\tilde\mu_{2j}$. Then, we normalize them by $\Sigma_N$ and we have the moments $M_j^{(N)}$ [Eq.~\eqref{normalized_moments_Mj}]. Before minimizing $\Delta$ with the L-BFGS-B method to obtain the Lagrange multipliers, we must rescale the problem. The minimization of $\Delta$ is carried out entirely with respect to the variable $\omega$. Once we get the rescaled multipliers $\lambda'_j$, we have that the rescaled density is:
\begin{equation}
H'^{}_N(\omega) = \exp\!\Big[-\lambda'^{}_0\,(1-\omega)^{N} - \sum_{j=1}^{N}\lambda'^{}_j\,B^{}_{N,j}(\omega)\Big]\,,
\label{eq:Hprime_sol}
\end{equation}
The rescaling is undone when reconstructing the transmission coefficient, as it needs to be expressed in the original variable $\tilde{k}$:
\begin{equation}
T^{}_N(\tilde{k})=\exp\left\lbrace-\frac{2\pi\Sigma^{}_N\,\tilde{k}}{(1+\tilde{k}^{2})^{N+2}}\,H^{}_N(w)\right\rbrace\,,
\label{eq:T_compact}
\end{equation}
with
\begin{equation}
H^{}_N(w)=\Sigma'^{}_N\,\rho^{2N+2}\,\left(\frac{1+\tilde{k}^{2}}{\tilde{k}^{2}+\rho^{2}}\right)^{N+2}\!\!\!\! H'^{}_N\left(\frac{\tilde{k}^{2}}{\tilde{k}^{2}+\rho^{2}}\right)\,,
\label{eq:H_undo}
\end{equation}
where
\begin{eqnarray}
H'^{}_N\left(\frac{\tilde{k}^{2}}{\tilde{k}^{2}+\rho^{2}}\right) & = &\exp\Bigl[ -\lambda'^{}_0\,\left(\frac{\rho^{2}}{\tilde{k}^{2}+\rho^{2}}\right)^{N} \nonumber\\[2mm]
& - & \sum_{j=1}^{N}\lambda'^{}_j\,B^{}_{N,j}\left(\frac{\tilde{k}^{2}}{\tilde{k}^{2}+\rho^{2}}\right) \Bigr]\,.
\end{eqnarray}
%

%%%%% FIGURE
\begin{figure}[ht!]
\centering
\includegraphics[width=0.49\textwidth]{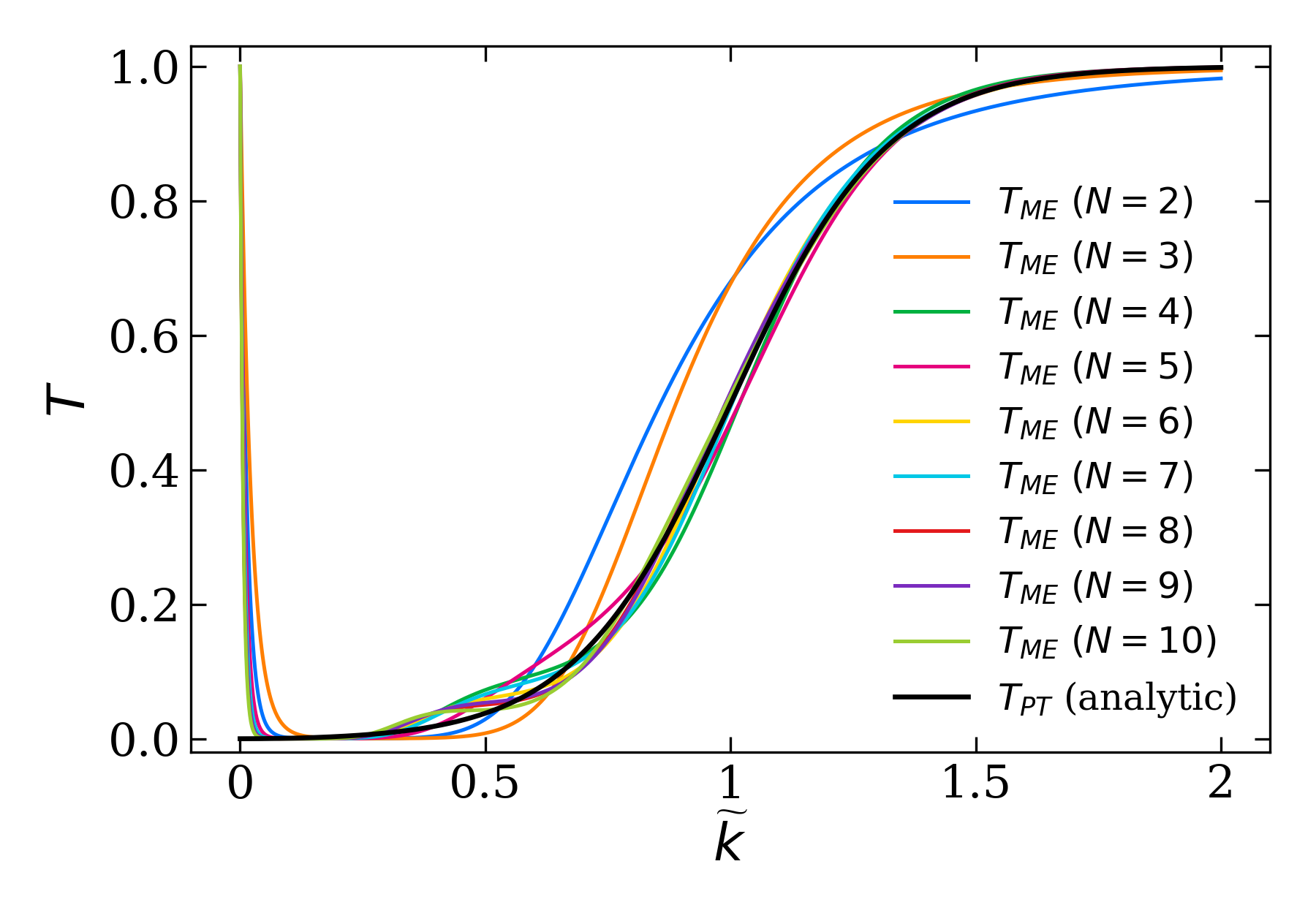} \\
\includegraphics[width=0.49\textwidth]{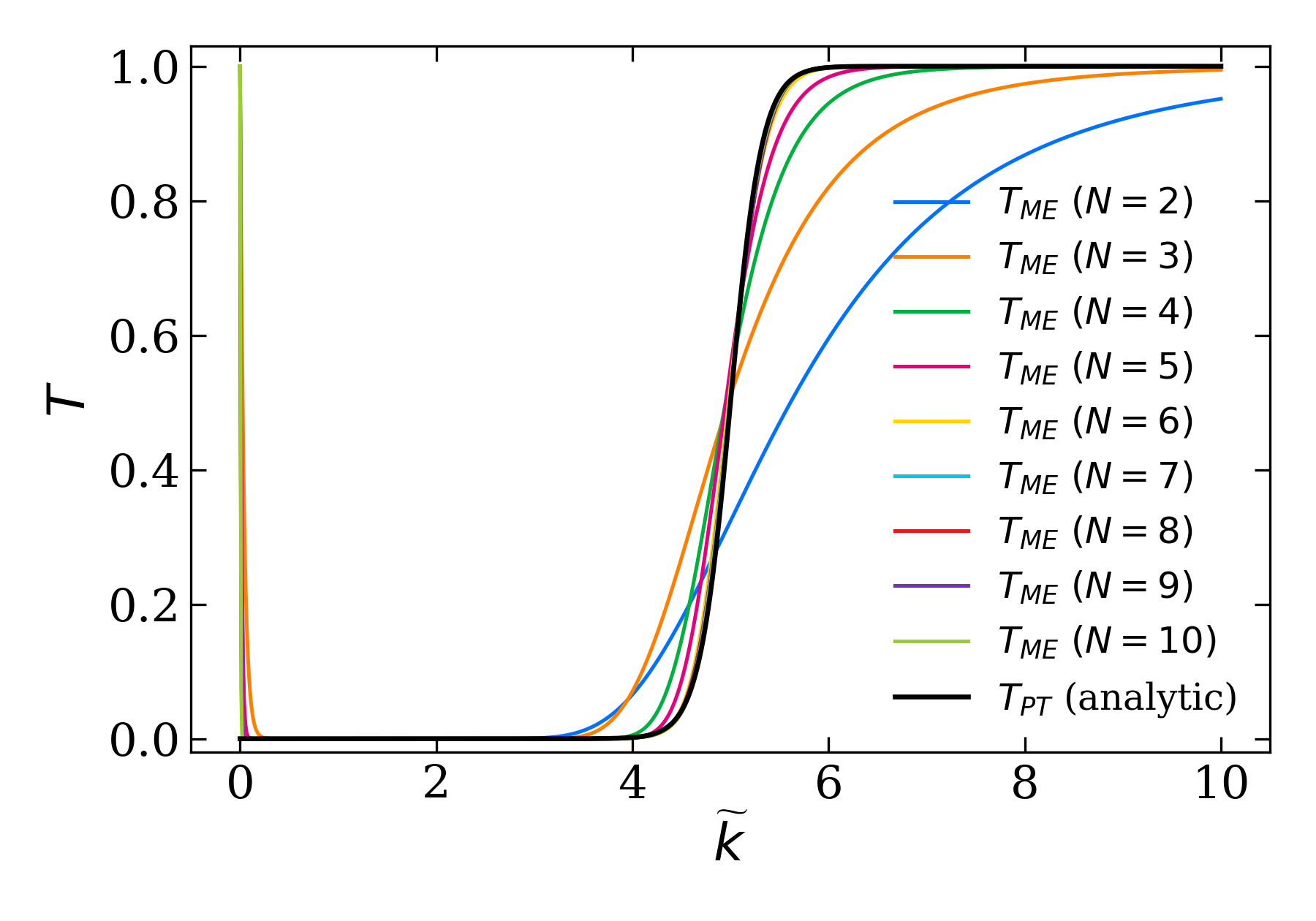} \\
\includegraphics[width=0.49\textwidth]{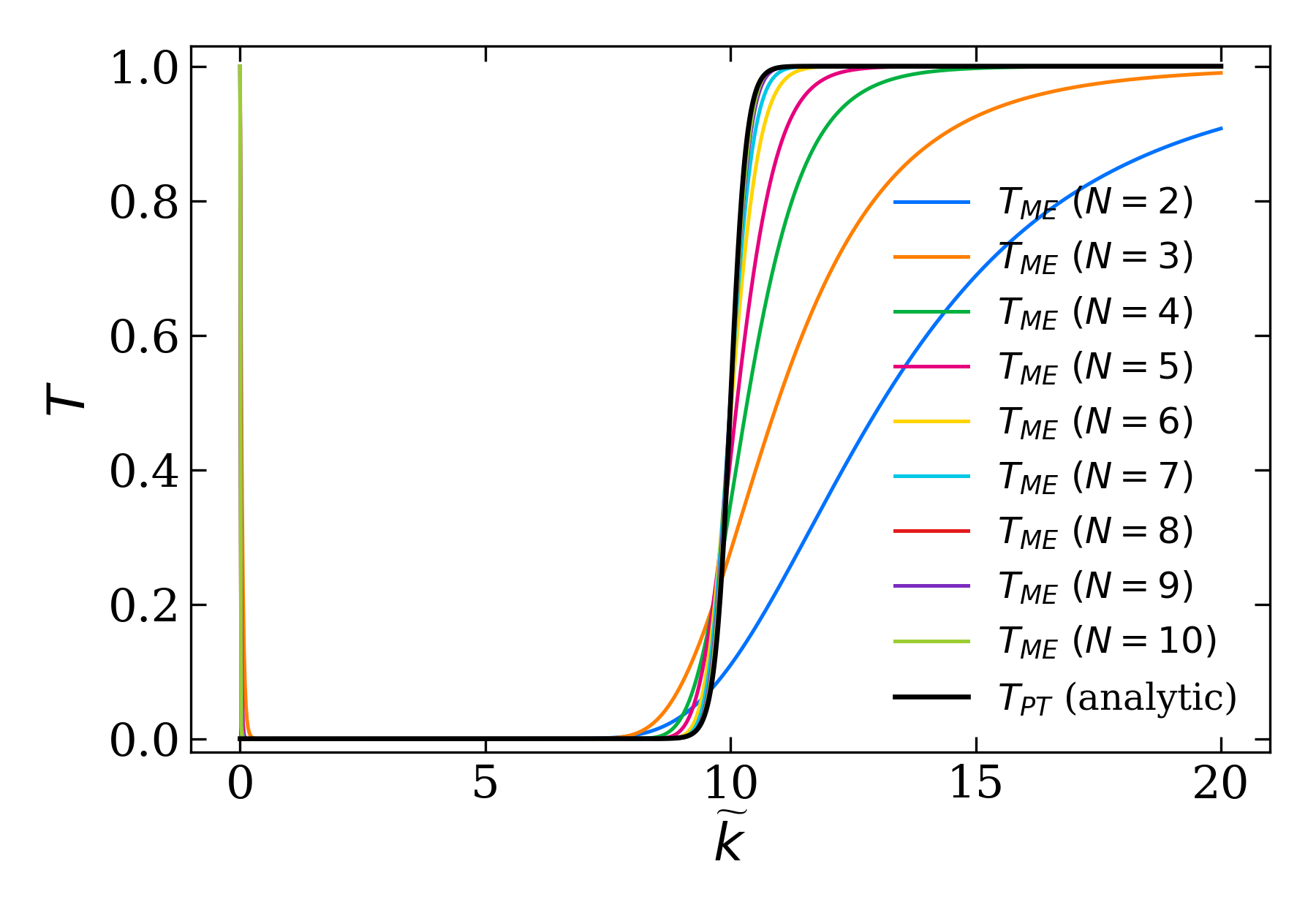}
\caption{Greybody factors for the P\"oschl-Teller potential as functions of the dimensionless frequency $\tilde{k}$, obtained from the MaxEnt reconstruction presented in Sec.~\ref{sec:compact} and compared with the exact solution for the following values of $\beta$: $\beta=1$ (top), $\beta=5$ (middle), and $\beta=10$ (bottom). The subscript $N$ indicates the number of moments used in the reconstruction.}
\label{fig:pt-reconstruction-compact}
\end{figure}
%%%%%

\Cref{fig:pt-reconstruction-compact} shows the reconstruction of $T_N(\tilde{k})$ for the P\"oschl--Teller potential with the compactified frequency domain, for the same values of $\beta$ as in~\Cref{fig:pt-reconstruction}. For every $\beta$ and $N$ the reconstruction breaks down at $\tilde{k}=0$, where $T_N$ jumps to $1$ instead of vanishing as the exact $T_{\rm PT}$ does. This is a direct consequence of the change of variables: after the compactification, the exponent in Eq.~\eqref{eq:T_compact} carries an overall factor of $\tilde{k}$, so that $T_N(0)=\exp(0)=1$ for any value of the multipliers. We call \emph{breakdown region} the narrow neighborhood of the origin in which $T_N$ departs from the (exponentially small) exact value, dropping from $T_N(0)=1$ over a finite interval. Defining its edge as the point at which $T_N$ falls below $10^{-2}$, we find that the region extends up to $\tilde{k}\simeq0.02$--$0.07$ for all $N$ except $N=3$, with only a weak dependence on $\beta$ and $N$. For $N=3$ it is wider, reaching $\tilde{k}\simeq 0.1$--$0.2$.

Away from the origin, the quality of the reconstruction depends strongly on $\beta$, which sets the position of the step, $\tilde{k}\sim\beta$. For $\beta=1$ the reconstruction with compactification is not the best option: the step is not well separated from the origin, and a small residual deviation remains at $\tilde{k}\lesssim0.7$ for all $N\geq4$. The reconstruction without compactification (\Cref{fig:pt-reconstruction}) reaches a better accuracy already with $N=2$--$4$. For larger values of $\beta$ ($\beta=5$ and $10$) the step lies far from the origin and the reconstruction improves systematically as $N$ increases.

The compactified formulation also removes several numerical ingredients required in Sec.~\ref{sec:maxent}: the integrals run over the finite interval $[0,1]$, and neither a clipping of the exponent nor bounds on the multipliers are needed. In the formulation without compactification, beyond a certain number of moments ($N\geq3$ for $\beta=1$, $N\geq7$ for $\beta=5$ and $N\geq6$ for $\beta=10$) the highest-order multiplier, which must be positive to guarantee the convergence of $Z$, ends up at its lower bound. The minimization then stops on that bound rather than at a stationary point, so the highest moments are not actually matched. This suggests that the non-monotonic behavior found there is mainly related to this constraint and to the numerical solution of the dual problem, and not to the information contained in the moments themselves. The truncation of the frequency domain and the clipping of the exponent may also contribute.
	
To quantify the accuracy of the reconstruction, we compare $T_N$ with the exact
Pöschl-Teller transmission probability, Eq.~(\ref{T_pt_real}). The
point-wise error,
\begin{equation}
  \delta T_{\rm PT}(\tilde{k}) = |T_N(\tilde{k}) - T_{\rm PT}(\tilde{k})|\,,
  \label{pointwise-error}
\end{equation}
measures the local deviation of the reconstruction with $N$ moments from the
exact result at each frequency. The integrated error,
\begin{equation}
  \Delta T_{\rm PT} = \int_{\tilde{k}_0}^{\tilde{k}_{\rm max}}
  \delta T_{\rm PT}(\tilde{k})\,d\tilde{k}\,,
  \label{integrated-error}
\end{equation}
is the area under $\delta T_{\rm PT}$ and summarises the accuracy over the whole
frequency range in a single number for each $(\beta,N)$. We take
$\tilde{k}_{\rm max}=2\beta$, so that the integration window contains the step
and the beginning of the plateau. For the reconstruction without
compactification $\tilde{k}_0=0$, while for the one with compactification
$\tilde{k}_0=0.15$, which excludes the breakdown region at the origin.

%%%%%
\begin{figure*}[ht!]
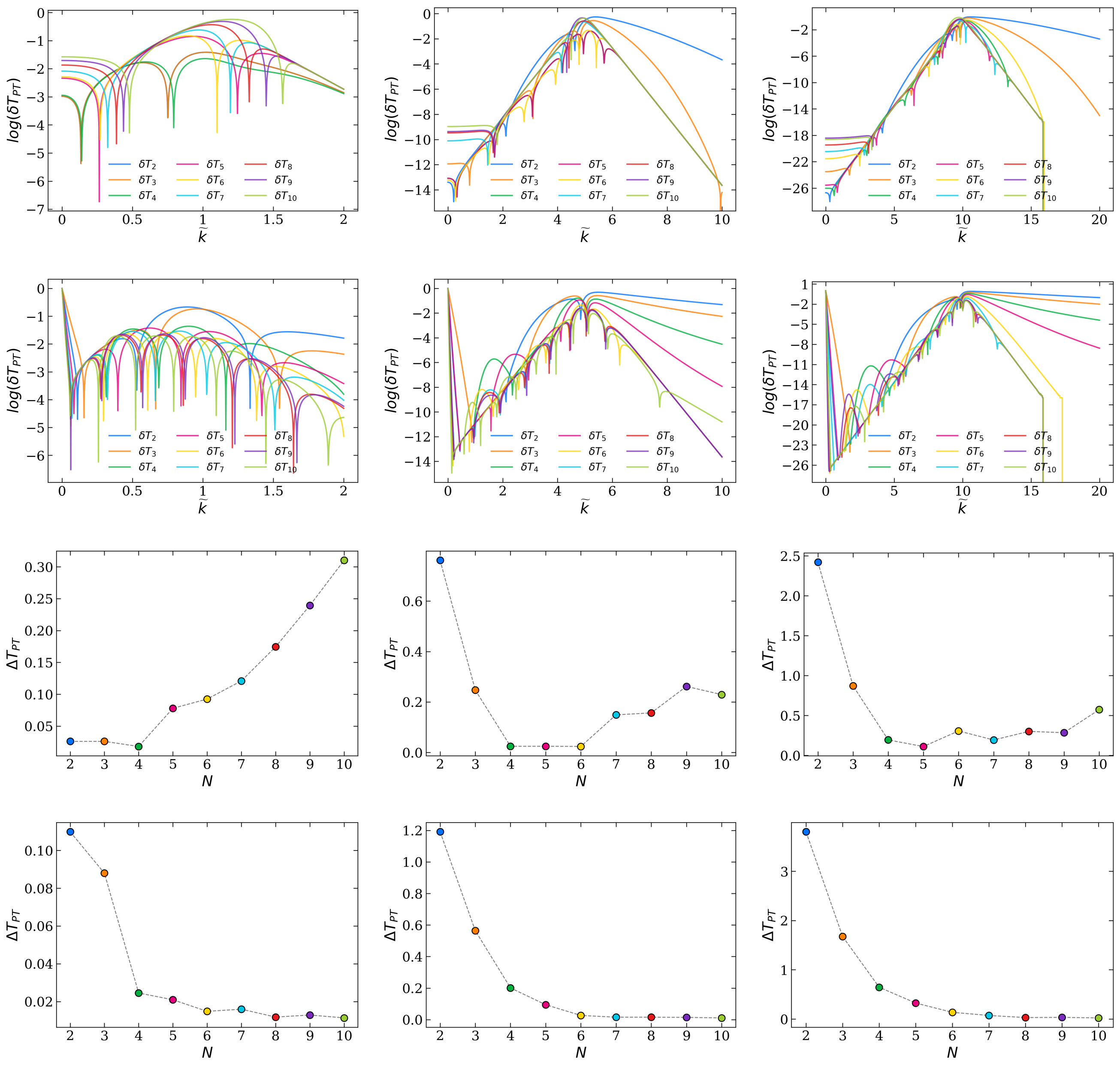

\centering
\newcommand{\bone}{1}
% <-- first beta: 1
%\newcommand{\figdir}{figures}
% folder with the PNGs
\newlength{\sfw}\setlength{\sfw}{0.325\textwidth}   % width of each mini figure
\newcommand{\sfig}[2]{\includegraphics[width=\sfw,height=0.7\sfw,keepaspectratio]{#1_beta_#2.png}}
\newcommand{\sfigc}[2]{\includegraphics[width=\sfw,height=0.7\sfw,keepaspectratio]{#1_beta_#2_compact.png}}
\setlength{\tabcolsep}{0pt}
\renewcommand{\arraystretch}{1.0}
\begin{tabular}{@{}ccc@{}}
\sfig{delta_T_PT}{\bone}  & \sfig{delta_T_PT}{5}  & \sfig{delta_T_PT}{10} \\
\sfigc{delta_T_PT}{\bone} & \sfigc{delta_T_PT}{5} & \sfigc{delta_T_PT}{10} \\
\sfig{N_PT}{\bone}        & \sfig{N_PT}{5}        & \sfig{N_PT}{10} \\
\sfigc{N_PT}{\bone}       & \sfigc{N_PT}{5}       & \sfigc{N_PT}{10} \\
\end{tabular}
\caption{Errors of the MaxEnt reconstruction of the P\"oschl-Teller greybody factor. Each column correspond to a different value of $\beta$: $\beta=1$ (left), $\beta=5$ (middle) and $\beta=10$ (right).
From top to bottom, the rows show: the point-wise error $\delta T_{\rm PT}(k)$ on a logarithmic scale of base $10$ for the standard reconstruction; and the same quantity for the reconstruction using the compactification.
The integrated error $\Delta T_{\rm PT}$ as a function of the number of moments $N$ for the standard reconstruction; and the same quantity for the reconstruction with compactification.}
\label{fig:pt-super}
\end{figure*}
%%%%%
	
\Cref{fig:pt-super} shows the point-wise error $\delta T_{\rm PT}$ (first two rows) and the integrated error $\Delta T_{\rm PT}$ (last two rows) with respect to the exact $T_{\rm PT}$. Every column corresponds to a value of $\beta$ (the ones we have already presented).

In the first row, we see the point-wise error for the standard reconstruction (without the compactification). The error is smallest for low $N$ and grows beyond it. For $\beta=1$, the best curve is $N=4$; for $N\geq5$ the error grows with $N$, both near the step and at small $\tilde{k}$. The same happens for $\beta=5$, where $N=6$ is best, and for $\beta=10$ with $N=5$. The integrated error in the third row confirms this: $\Delta T_{\rm PT}$ has a minimum at $N=4$ ($\beta=1$), $N=6$ ($\beta=5$, although $N=4,5,6$ are indistinguishable at the scale of the figure) and $N=5$ ($\beta=10$), and then grows, although not monotonically (for $\beta=5$ the value at $N=10$ is slightly below the one at $N=9$, and for $\beta=10$ the error at $N=7$ is comparable to the one at $N=4$).
In the second row, we show the point-wise error for the reconstruction using the compactification. Here, $\delta T_{\rm PT}(0)=1$ for all $N$ and $\beta$, but the error then drops by many orders of magnitude, since both $T_N$ and $T_{\rm PT}$ are exponentially small there. Away from the origin the trend is the opposite to the case without compactification: the error decreases with $N$. We see in the fourth row that $\Delta T_{\rm PT}$ decreases, up to small non-monotonic bumps. Since the error decreases essentially monotonically, $N$ can be chosen just by checking the convergence of the reconstruction, without knowing the exact result.

%%%%% FIGURE
\begin{figure*}[ht!]
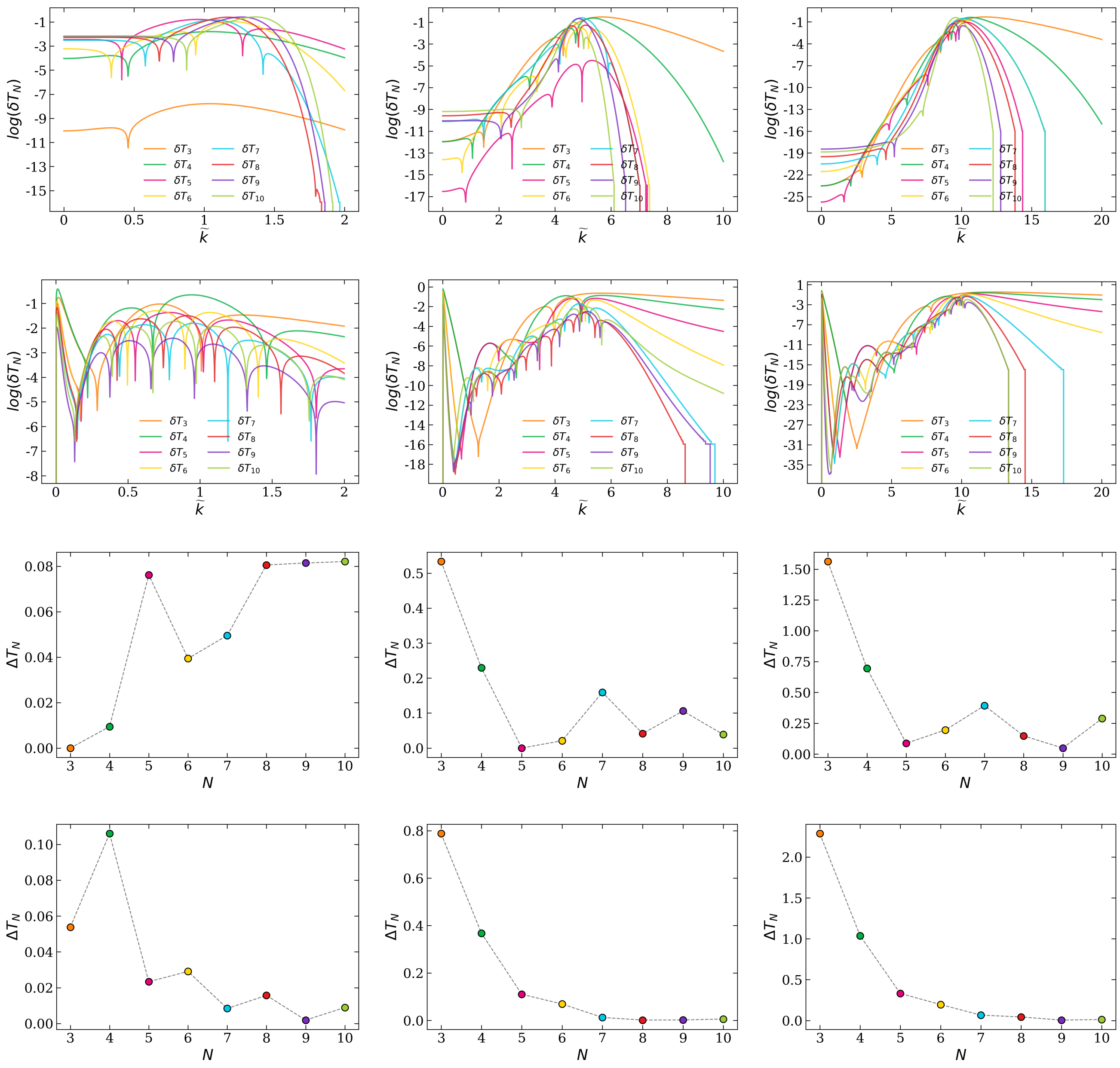

\centering
\newcommand{\bone}{1}
% <-- first beta: 1
%\newcommand{\figdir}{figures}
% folder with the PNGs
\setlength{\sfw}{0.325\textwidth}   % width of each mini figure
\newcommand{\sfig}[2]{\includegraphics[width=\sfw,height=0.7\sfw,keepaspectratio]{#1_beta_#2.png}}
\newcommand{\sfigc}[2]{\includegraphics[width=\sfw,height=0.7\sfw,keepaspectratio]{#1_beta_#2_compact.png}}
\setlength{\tabcolsep}{0pt}
\renewcommand{\arraystretch}{1.0}
\begin{tabular}{@{}ccc@{}}
\sfig{delta_T_N}{\bone}  & \sfig{delta_T_N}{5}  & \sfig{delta_T_N}{10} \\
\sfigc{delta_T_N}{\bone} & \sfigc{delta_T_N}{5} & \sfigc{delta_T_N}{10} \\
\sfig{N_N}{\bone}        & \sfig{N_N}{5}        & \sfig{N_N}{10} \\
\sfigc{N_N}{\bone}       & \sfigc{N_N}{5}       & \sfigc{N_N}{10} \\
\end{tabular}
\caption{Errors between consecutive MaxEnt reconstructions. Columns correspond to the following values of $\beta$: $\beta=1$ (left), $\beta=5$ (middle) and $\beta=10$ (right).
The rows show, from top to bottom, the successive difference between reconstructions $\delta T_{N}$ on a logarithmic scale of base 10 for the standard reconstruction; the same quantity for the reconstruction using the compactification;
the integrated version $\Delta T_{N}$ as a function of the number of moments $N$ for the standard reconstruction; and the same quantity for the reconstruction using the compactification.}
\label{fig:pt-super-2}
\end{figure*}
%%%%%

A complementary diagnostic, which is useful when no exact solution is available, as in the case of the BH potentials, is the successive difference between reconstructions,
\begin{equation}
\delta T^{}_N(\tilde{k}) = |T^{}_N(\tilde{k}) - T^{}_{N-1}(\tilde{k})|\,,
\label{successive-error}
\end{equation}
together with its integrated version,
\begin{equation}
\Delta T^{}_N = \int_{\tilde{k}^{}_0}^{\tilde{k}^{}_{\max}} \delta T^{}_N(\tilde{k})\,d\tilde{k}\,,
\label{integrated-successive-error}
\end{equation}
with the same $\tilde{k}_0$ and $\tilde{k}_{\max}$ as in Eq.~\eqref{integrated-error}.

\Cref{fig:pt-super-2} shows both quantities for $\beta=1,5,10$ in the same way as in \Cref{fig:pt-super}. The successive differences $\delta T^{}_N$ from Eq.~\eqref{successive-error}, in the first row, do not decay with $N$, so the sequence $T_N$ does not converge. The integrated differences $\Delta T^{}_N$ from Eq.~\eqref{integrated-successive-error}, in the third row, give the same picture: for $\beta=1$ they grow and saturate at $\Delta T_N\simeq0.08$, while for $\beta=5$ and $10$ they decrease up to $N=5$ and then fluctuate without any systematic decay.

Only two cases show an almost vanishing $\Delta T_N$: $N=3$ for $\beta=1$ ($\Delta T_3\sim10^{-10}$) and $N=5$ for $\beta=5$ ($\Delta T_5\sim10^{-5}$). By construction of Eq.~\eqref{integrated-successive-error}, this means that $T_3\simeq T_2$ and $T_5\simeq T_4$, respectively, which is why these curves overlap in \Cref{fig:pt-reconstruction}. Such coincidences are not evidence of convergence: a small successive difference is necessary but not sufficient, since it only indicates that the added moment barely modifies the reconstruction. The decisive quantity is the distance to the exact result, $\Delta T_{\rm PT}$, shown in the third row of \Cref{fig:pt-super}. After an initial decrease at low $N$, it does not decrease any further: for $\beta=1$ it grows monotonically for $N\geq4$, and for $\beta=5$ and $10$ it grows or fluctuates beyond $N=5$. Adding moments therefore does not bring $T_N$ closer to the exact greybody factor.

Regarding the reconstruction using the compactification, in the second row of~\Cref{fig:pt-super-2} we can see that $\delta T_N(0)=0$ for all $N$, since all $T_N$ share the same value at the origin, where the approach that uses the compactification breaks down. Immediately after, $\delta T_N$ shows a narrow peak and then drops by many orders of magnitude. Excepting for the origin, and in contrast to the standard reconstruction, $\delta T_N$ decreases with $N$. Likewise, $\Delta T_N$ in the fourth row decreases with $N$, with only minor non-monotonic fluctuations (except for $\beta=1$, where they are more pronounced). Since adding moments no longer changes the reconstruction, we stop at $N=10$.

\Cref{tab:optimal-N} compares, for the formulations with and without compactification, the minimum over $N$ of the integrated error with respect to the exact result. We quantify the improvement as
\begin{equation}
\eta\, [\%] = \frac{\min \Delta T^{\mathrm{NC}}_{\mathrm{PT}} - \min \Delta T^{\mathrm{C}}_{\mathrm{PT}}}{\min \Delta T^{\mathrm{NC}}_{\mathrm{PT}}}\times 100\% \,.
\label{improvement}
\end{equation}
%

%%%%% TABLE
\begin{table}[h]
\centering
\begin{tabular}{c c c c c c}
\cline{2-5}
& \multicolumn{2}{|c|}{{Reconstruction}} & \multicolumn{2}{c|}{{Reconstruction}} & \\
& \multicolumn{2}{|c|}{without} & \multicolumn{2}{c|}{with} &  \\
& \multicolumn{2}{|c|}{{compactification}} & \multicolumn{2}{c|}{{compactification}} &  \\
\hline
\multicolumn{1}{|c|}{$\beta$} & \multicolumn{1}{|c|}{$N^{}_{\rm opt}$} & \multicolumn{1}{c|}{$\min\Delta T^{}_{\rm PT}$} & \multicolumn{1}{|c|}{$N^{}_{\rm opt}$} & \multicolumn{1}{c|}{$\min\Delta T^{}_{\rm PT}$} & \multicolumn{1}{|c|}{\% of Improvement} \\
\hline
\multicolumn{1}{|c|}{1}  & \multicolumn{1}{|c|}{4} & \multicolumn{1}{c|}{$1.82 \times 10^{-2}$} & \multicolumn{1}{|c|}{10} & \multicolumn{1}{c|}{$1.14 \times 10^{-2}$} & \multicolumn{1}{|c|}{$37.53\%$} \\
\hline
\multicolumn{1}{|c|}{5}  & \multicolumn{1}{|c|}{6} & \multicolumn{1}{c|}{$2.38 \times 10^{-2}$} & \multicolumn{1}{|c|}{10} & \multicolumn{1}{c|}{$1.09 \times 10^{-2}$} & \multicolumn{1}{|c|}{$54.38\%$} \\
\hline
\multicolumn{1}{|c|}{10} & \multicolumn{1}{|c|}{5} & \multicolumn{1}{c|}{$1.12 \times 10^{-1}$} & \multicolumn{1}{|c|}{10} & \multicolumn{1}{c|}{$2.26 \times 10^{-2}$} & \multicolumn{1}{|c|}{$79.81\%$} \\
\hline
\end{tabular}
\caption{Minimum integrated error $\Delta T_{\rm PT}$ associated with the plots shown~\Cref{fig:pt-super} for the MaxEnt reconstruction without (third row) and with (fourth row) compactification. The improvement is calculated using Eq.~\eqref{improvement}.}
\label{tab:optimal-N}
\end{table}
%%%%%

For $\beta=5$ and $\beta=10$, the integrated error of the best compactified reconstruction is $54\%$ and $80\%$ lower, respectively, than that of the best standard one. In addition, in the compactified formulation the error decreases systematically with $N$, so a suitable $N$ can be chosen without knowing the exact result. In the standard formulation, by contrast, the error is non-monotonic in $N$, and the optimal value can only be identified by comparing with $T_{\rm PT}$, which is not available in practice. This supports the use of the compactification for these values of $\beta$.

For $\beta=1$ the situation is different. The compactified reconstruction with $N=10$ still gives a lower integrated error, but the gain is smaller ($38\%$). The region $\tilde{k}<0.15$, which is excluded from the integral for all values of $\beta$, is not negligible here: the transition of $T_{\rm PT}$ lies close to the origin ($\tilde{k}\sim1$), so the breakdown region and the spurious oscillations of $\delta T_{\rm PT}$ affect a sizeable part of the relevant range, whereas for $\beta=5$ and $10$ the transition is far from the origin. Since the improvement relies on discarding a region that matters for $\beta=1$, we prefer the standard reconstruction with $N=4$ in this case.	
	
These results demonstrate that the MaxEnt method, applied to a modest number of KdV integrals, reconstructs the P\"oschl-Teller greybody factor with good accuracy. The form without compactification is the best option when only a few moments are available or when $\beta$ is small, but its accuracy deteriorates when more moments are added and it depends on a number of ad hoc truncations. The form with compactification is clean and converges systematically for large $\beta$, but it requires more moments and it cannot reproduce the region around $\tilde{k}=0$. This behaviour will also be reflected in the following section, when we apply the method to the BH potentials, where no exact solution is available.

%%%%% GREYBODY FACTORS FOR THE SCHWARZSCHILD BH
\section{Greybody Factors for the Schwarzschild BH}
\label{sec:rw}
	
We now apply the MaxEnt method to the Regge-Wheeler BH potential in Eq.~\eqref{RW-potential}, which describes the response of the BH to axial perturbations of a Schwarzschild BH. Unlike the P\"oschl-Teller case, no closed-form expression for the greybody factor are available nor expected. The accuracy of the MaxEnt reconstruction is assessed by comparison with results of the WKB approximation~\cite{Schutz:1985km,Iyer:1986np,Iyer:1986nq}.
	
Using a dimensionless radial variable, $\tilde{r} = r/ r_s$, the Regge-Wheeler potential takes the simpler form
\begin{equation}
V_{\ell}^{\rm RW}(\tilde{r}) = \left(1 - \frac{1}{\tilde{r}}\right)\left(\frac{\ell(\ell+1)}{\tilde{r}^2} - \frac{3}{\tilde{r}^3}\right)\,.
\label{RW-dimensionless}
\end{equation}
The expressions for the first KdV integrals for this BH potential are given in Appendix~\ref{app:kdv-rw}.

\Cref{fig:rw-reconstruction} compares the MaxEnt reconstructions with the fourth-order WKB approximation for $\ell=2,5,10$. For the compactified formulation we take $N=10$ in all cases, since the error decreases approximately monotonically with $N$, as found for the P\"oschl--Teller potential. In the standard formulation the behavior is non-monotonic in $N$, so the optimal choice is not obvious a priori. For transitions far from the origin, $N=5$ is already a good choice, as in the P\"oschl--Teller benchmark for $\beta=5$ and $10$. The case $\ell=2$ requires more care: its transition lies close to the origin, as for $\beta=1$, where the benchmark favors a smaller $N$ ($N=4$) and the standard error worsens as moments are added. However, the reconstructions with $N=4$ and $N=5$ turn out to be practically identical: their integrated difference, computed excluding the region $\tilde k<0.15$ as in the P\"oschl--Teller case, is of order $10^{-8}$. We therefore use $N=5$ for $\ell=2$ as well, as for the other multipoles.

%%%%%
\begin{figure}[ht!]
\centering
\includegraphics[width=0.48\textwidth]{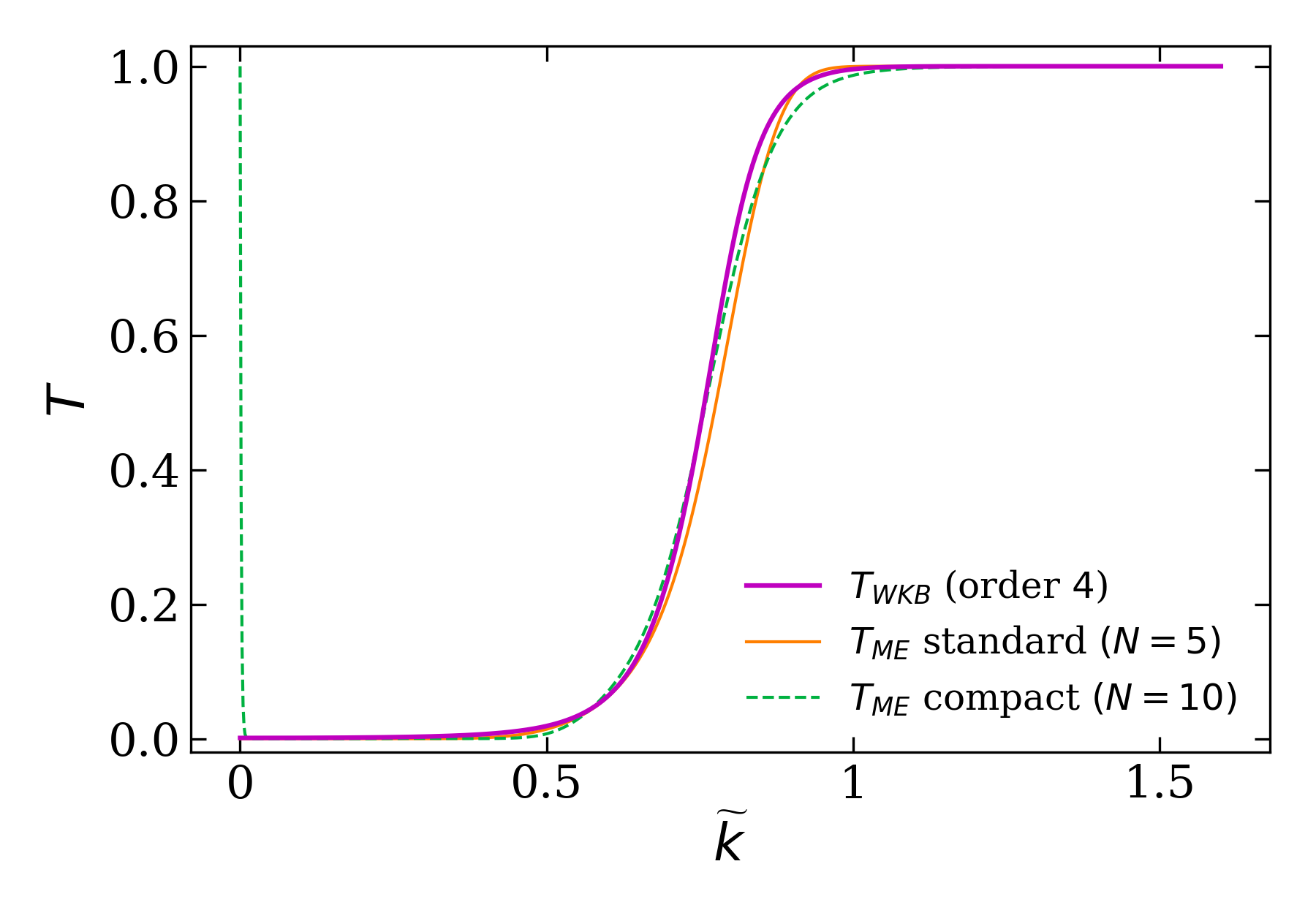} \\
\includegraphics[width=0.48\textwidth]{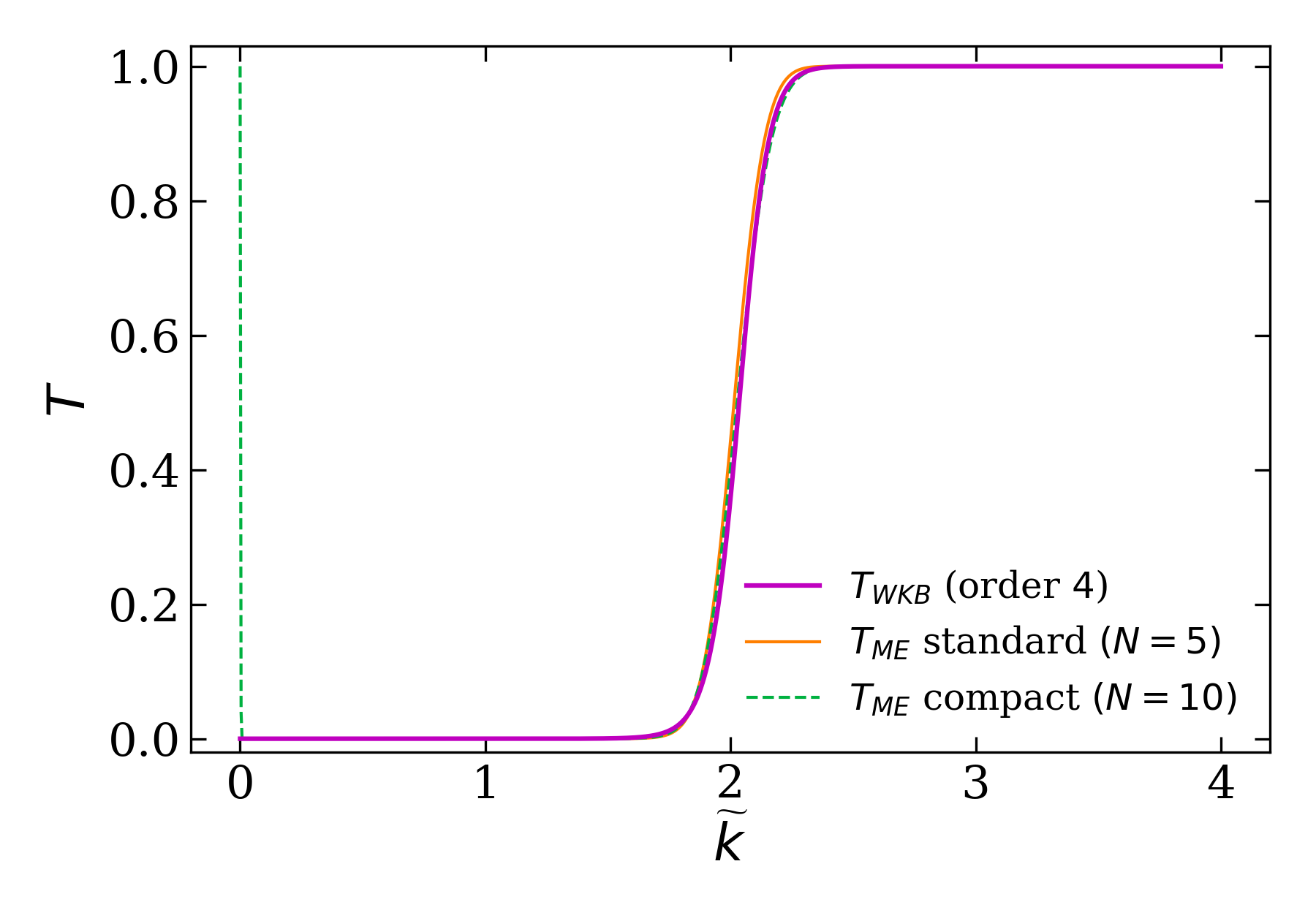} \\
\includegraphics[width=0.48\textwidth]{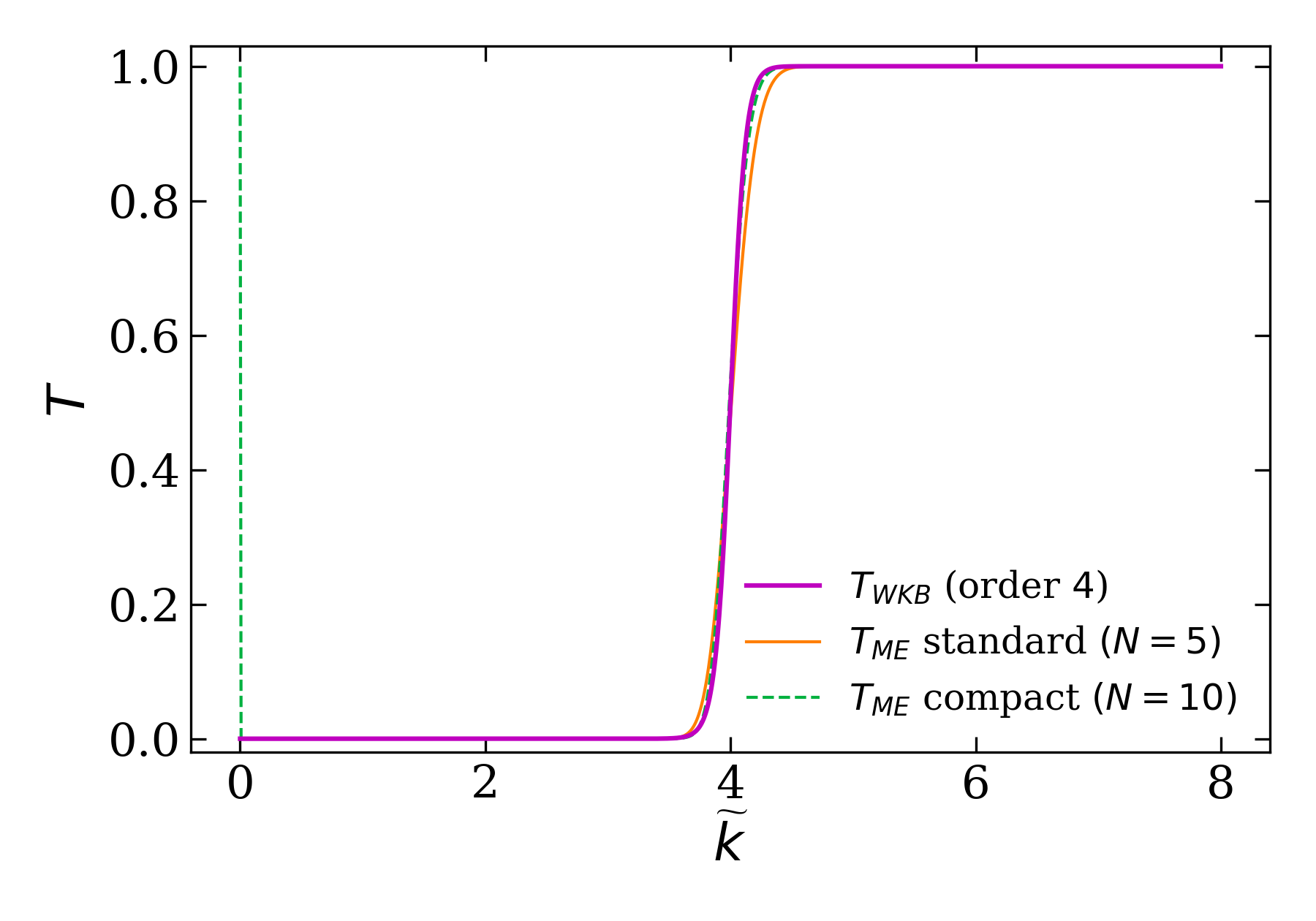}
\caption{Greybody factors for the Regge-Wheeler potential as functions of the dimensionless frequency $\tilde{k}$ for $\ell=2$ (top), $\ell=5$ (middle) and $\ell=10$ (bottom). The results of the MaxEnt reconstructions without compactification ($N=5$, orange) and with compactification ($N=10$, green, dashed) are compared with the fourth-order WKB approximation.}
\label{fig:rw-reconstruction}
\end{figure}
%%%%%

For $\ell=5$ and $\ell=10$ the step lies far from the origin, and the compactified curves are nearly indistinguishable from WKB at the scale of the figure, describing the central part of the transition better than the standard ones. For $\ell=2$ neither reconstruction is uniformly closer to WKB. The standard reconstruction coincides with WKB at small $\tilde k$ and for $\tilde k\gtrsim1$, but is shifted towards larger $\tilde k$ in the transition region. The compactified one follows WKB more closely in the rising part of the transition, but it slightly underestimates $T$ for $\tilde k\simeq0.40$--$0.55$ and approaches $T=1$ more slowly; its integrated difference with respect to WKB is only somewhat lower ($19\%$) than that of the standard one. 

In all cases the compactified curves jump to $T=1$ at $\tilde k=0$, as discussed in Sec.~\ref{sec:compact}. Since WKB is itself an approximation, the residual differences cannot be attributed solely to the reconstruction.

It is worth remembering that by virtue of the Darboux covariance, these results extend to any admissible BH potential, in particular for the Zerilli potential of the polar sector of the Schwarzschild metric perturbations.
	
The MaxEnt method offers several advantages over other approaches. Unlike Pad\'e approximants~\cite{Lenzi:2023inn}, which introduce poles that require physical interpretation, the MaxEnt reconstruction is an entire function of $k$ with no spurious singularities. Unlike the WKB approximation, which becomes increasingly complicated at higher orders, the MaxEnt method can be systematically improved by including additional moments within the limits set by the convergence of the dual problem, with the structure of the calculation remaining unchanged.

%%%%%
\section{Conclusions and Discussion}
\label{sec:conclusions}
	
In this paper, we have developed and validated a new method for computing BH greybody factors using the MaxEnt principle. 
The method builds on recent work showing that the KdV integrals of the BH potential fully determine the greybody factors via a classical moment problem~\cite{Lenzi:2022wjv,Lenzi:2021njy,Lenzi:2021wpc}. The MaxEnt method selects the least-biased distribution compatible with a finite set of moments, guaranteeing positivity at every truncation order~\cite{Meadpapanicolau1984}.
	
We first validated the method against the exactly solvable P\"oschl-Teller potential, where we used two formulations of the moment problem: the original one on the semi-infinite domain and a reformulation that compactifies the frequency domain onto $[0,1]$, leaving a moment problem similar to a Hausdorff moment problem but not identical.  
The resulting method, instead of using moments defined with the powers of the frequency, $k^n$, uses the Bernstein polynomials, which appear in a natural way from the way in which we carry out the compactification. 
The two formulations are complementary. The standard one is the best option for small $\beta$ or when only a few moments are available: its integrated error with respect to the exact result reaches a minimum at $N_{\rm opt}=4$, $6$ and $5$ for $\beta=1$, $5$ and $10$, respectively (\Cref{tab:optimal-N}), and grows beyond that point. Its accuracy, however, deteriorates when more moments are added. 
Our results also suggest that the standard method is limited by the numerical solution of the dual problem: starting at $N=3$ for $\beta=1$, $N=7$ for $\beta=5$ and $N=6$ for $\beta=10$, the highest-order multiplier, which must be positive, ends up at its lower bound. The minimization then stops on that bound instead of at a stationary point, so the highest moments are not matched. The truncation of the infinite frequency domain and the clipping of the exponent may also contribute.
The formulation that uses the compactification converges systematically with $N$ for large $\beta$. The price is that it cannot reproduce a tiny region around the origin of the frequency, where $T_N$ jumps to $1$ for any value of the Lagrange multipliers used in the MaxEnt method.  This appears to be a structural consequence of the compactification, not a failure of the minimization procedure.
	
We then applied the method to the Regge--Wheeler potential governing axial perturbations of a Schwarzschild BH, for $\ell=2,5,10$. The compactified reconstruction with $N=10$ agrees closely with the fourth-order WKB approximation away from the breakdown region, while the standard one with $N=5$ agrees somewhat less well. For $\ell=2$, whose transition lies close to the origin, neither formulation is uniformly closer to WKB. 

This is the first application of the MaxEnt method to the reconstruction of BH greybody factors, and it demonstrates that a modest number of KdV integrals is sufficient to recover them accurately.
	
The MaxEnt method has several advantages for this problem: (i) it guarantees the positivity of the
reconstructed distribution at every truncation order; (ii) the reconstruction is an entire function of
$k$, with no spurious poles; (iii) additional moments can be incorporated in a straightforward way; and
(iv) the dual problem is convex, so standard optimizers like L-BFGS-B can be used, although convergence at high $N$ requires care.

The method can be extended to other situations of physical interest, such as (i) perturbations of
rotating (Kerr) BHs, using the Teukolsky formalism~\cite{Teukolsky:1972le,Teukolsky:1973ha}; (ii)
perturbations of exotic compact objects that mimic BHs; (iii) perturbations of BHs in alternative
theories of gravity; (iv) perturbations of BHs in higher dimensions; and (v) scalar, vector, and
neutrino perturbations, in addition to gravitational ones.
	
The connection between KdV integrals, DTs, and greybody factors may also provide new insights into the structure of BH perturbation theory and its relationship to integrable systems. The fact that the KdV integrals are invariant under DTs suggests a deep connection between the isospectral symmetries of the Schr\"odinger equation and the physical scattering data.
	
In summary, the MaxEnt method provides a systematic and computationally inexpensive route to BH greybody factors from KdV integrals, with positivity guaranteed at every truncation order. Its accuracy is, however, limited by the numerical solution of the dual problem. In the semi-infinite formulation the highest-order multiplier reaches its lower bound from $N=3$ ($\beta=1$), $N=7$ ($\beta=5$) and $N=6$ ($\beta=10$); the minimization then stops on the bound and the highest moments are not matched. The compactified formulation remains stable up to the highest order considered ($N=10$), but it cannot reproduce a narrow region around the origin, whose width depends on $\beta$ and $N$ ($\tilde k\lesssim0.02$--$0.20$ in our runs), and its error decreases with $N$ only up to small non-monotonic fluctuations. In practice, a modest number of moments, chosen according to the formulation and the potential, yields accurate reconstructions.

%%%%% ACKNOWLEDMENTS
\begin{acknowledgments}
CFS is supported by contracts PID2019-106515GB-I00/AEI/10.13039/501100011033 and PID2025-173920NB-C21/AEI/10.13039/501100011033 (Spanish Ministry of Science and Innovation) and 2025 SGR 00191 (AGAUR, Generalitat de Catalunya). 
The work of ML is supported by the F.R.S.-FNRS Belgium through convention IISN 4.4516.26 and CDR grant J.0006.26.
This work was also partially supported by the program Unidad de Excelencia Mar\'{\i}a de Maeztu CEX2020-001058-M (Spanish Ministry of Science and Innovation).
We have used the Computer Algebra System Maxima~\cite{maxima} and the SciPy library~\cite{Virtanen2020} for the numerical computations in this paper.
\end{acknowledgments}

%%%%% APPENDICES
\appendix

%%%%% KDV INTEGRALS FOR THE POSCHL-TELLER POTENTIAL
\section{KdV Integrals for the P\"oschl-Teller Potential}
\label{app:kdv-pt}
	
For the reader's convenience, we reproduce here the first non-vanishing KdV integrals of
the P\"oschl-Teller potential~\eqref{PT-potential}, originally given in Appendix~B of
Ref.~\cite{Lenzi:2025kqs}, so that the paper is self-contained. In dimensionless form
(with $\alpha=1$), they read:
\begin{widetext}
\begin{eqnarray}
\mathcal{K}^{}_1 & = & \frac{1}{2}\left(4\beta^2 + 1\right)\,, \\[2mm]
\mathcal{K}^{}_3 & = & -\frac{1}{12}\left(4\beta^2 + 1\right)^2\,, \\[2mm]
\mathcal{K}^{}_5 & = & \frac{1}{30}\left(4\beta^2 + 1\right)^2\left(4\beta^2 + 3\right)\,, \\[2mm]
\mathcal{K}^{}_7 & = & -\frac{1}{168}\left(4\beta^2 + 1\right)^2\left(48\beta^4 + 88\beta^2 + 51\right)\,, \\[2mm]
\mathcal{K}^{}_9 & = & \frac{1}{90}\left(4\beta^2 + 1\right)^2\left(64\beta^6 + 208\beta^4 + 284\beta^2 + 155\right)\,, \\[2mm]
\mathcal{K}^{}_{11} & = & -\frac{1}{132}\left(4\beta^2 + 1\right)^2\left(256\beta^8 + 1280\beta^6 + 3040\beta^4 + 3856\beta^2 + 2073\right)\,, \\[2mm]
\mathcal{K}^{}_{13} & = & \frac{1}{2730}\left(4\beta^2 + 1\right)^2\left(15360\beta^{10} + 108800\beta^8 + 393088\beta^6 + 860576\beta^4 + 1070668\beta^2 \right. \nonumber \\[2mm]
& & \left. + 573405\right)\,, \\[2mm]
\mathcal{K}^{}_{15} & = & -\frac{1}{720}\left(4\beta^2 + 1\right)^2\left(12288\beta^{12} + 116736\beta^{10} + 593152\beta^8 + 1965824\beta^6 + 4210384\beta^4 + 5212104\beta^2 \right. \nonumber \\[2mm]
& & \left. + 2788707\right)\,, \\[2mm]
\mathcal{K}^{}_{17} & = & \frac{1}{1530}\left(4\beta^2 + 1\right)^2\left(81920\beta^{14} + 1003520\beta^{12} + 6804480\beta^{10} + 31612160\beta^8 + 102323648\beta^6 \right. \nonumber \\[2mm]
& & \left. + 217896816\beta^4 + 269394108\beta^2 + 144103095\right)\,, \\[2mm]
\mathcal{K}^{}_{19} & = & -\frac{1}{7980}\left(4\beta^2 + 1\right)^2\left(1376256\beta^{16} + 21102592\beta^{14} + 183844864\beta^{12} + 1137156096\beta^{10} \right. \nonumber \\[2mm]
& & + 5153682944\beta^8 + 16577459712\beta^6 + 35249267776\beta^4 + 43565974240\beta^2 \nonumber \\[2mm]
& & \left. + 23302711005\right)\,.
\end{eqnarray}
\end{widetext}

%%%%% KDV INTEGRALS FOR THE REGGE-WHEELER POTENTIAL
\section{KdV Integrals for the Regge-Wheeler Potential}
\label{app:kdv-rw}
	
For the reader's convenience, we reproduce here the first non-vanishing KdV integrals of
the Regge-Wheeler potential~\eqref{RW-potential}, originally computed in Appendix E of Ref.~\cite{Lenzi:2023inn}, so that the paper is self-contained. They are given in units
$r_s=1$, in terms of $L=\ell(\ell+1)$:
\begin{widetext}
\begin{eqnarray}
\mathcal{K}^{}_1 & = & L - \frac{3}{2}\,, \\[2mm]
\mathcal{K}^{}_3 & = & -\left(\frac{1}{12}L^2 - \frac{3}{10}L + \frac{3}{10}\right)\,, \\[2mm]
\mathcal{K}^{}_5 & = & \frac{2}{105}L^3 - \frac{83}{840}L^2 + \frac{5}{28}L - \frac{29}{280}\,, \\[2mm]
\mathcal{K}^{}_7 & = & -\left(\frac{1}{168}L^4 - \frac{53}{1386}L^3 + \frac{263}{2772}L^2 - \frac{147}{1430}L + \frac{222}{5005}\right)\,, \\[2mm]
\mathcal{K}^{}_9 & = & \frac{14}{6435}L^5 - \frac{41}{2574}L^4 + \frac{8}{165}L^3 - \frac{2557}{34320}L^2 + \frac{1203}{19448}L - \frac{723}{38896}\,, \\[2mm]
\mathcal{K}^{}_{11} & = & -\left(\frac{1}{1144}L^6 - \frac{67}{9724}L^5 + \frac{73}{2992}L^4 - \frac{186859}{3879876}L^3 + \frac{2253239}{38798760}L^2 - \frac{19129}{587860}L + \frac{64759}{5173168}\right)\,,\\[2mm]
\mathcal{K}^{}_{13} & = & \frac{11}{29393}L^7 - \frac{1529}{503880}L^6 + \frac{3091}{251940}L^5 - \frac{103699}{3527160}L^4 + \frac{2719789}{60843510}L^3 - \frac{8348939}{243374040}L^2 + \frac{39}{1292}L\nonumber \\[2mm]
& &  + \frac{4803}{772616}\,, \\[2mm]
\mathcal{K}^{}_{15} & = & -\left(\frac{13}{77520}L^8 - \frac{13}{9690}L^7 + \frac{91}{14535}L^6 - \frac{28639}{1671525}L^5 + \frac{81283}{2674440}L^4 - \frac{7109}{222870}L^3 + \frac{111529}{2228700}L^2\right. \nonumber \\[2mm]
& & \left. + \frac{17637}{566950}L + \frac{1194813}{21544100}\right)\,, \\[2mm]
\mathcal{K}^{}_{17} & = & \frac{26}{334305}L^9 - \frac{131}{222870}L^8 + \frac{16}{4845}L^7 - \frac{217519}{23401350}L^6 + \frac{750447}{37702175}L^5 - \frac{1106101}{37702175}L^4 \nonumber \\[2mm]
& & + \frac{83799194}{1676927175}L^3 + \frac{18613463771}{176316914400}L^2 + \frac{1930951511}{7260108240}L + \frac{83062407817}{246843680160}\,, \\[2mm]
\mathcal{K}^{}_{19} & = & -\left(\frac{17}{458850}L^{10} + \frac{1003}{3991995}L^9 + \frac{48841}{26613300}L^8 - \frac{886822}{206253075}L^7 + \frac{148953983}{9900147600}L^6\right. \nonumber \\[2mm]
& & - \frac{1144342879}{54450811800}L^5 + \frac{600058031}{24200360800}L^4 + \frac{19712925427}{163352435400}L^3 + \frac{290243073581}{326704870800}L^2\nonumber \\[2mm]
& & \left. - \frac{43168543419}{21880401400}L + \frac{4524494561511}{1662910506400}\right)\,.
\end{eqnarray}
\end{widetext}
%

%%%%% APPENDIX: DETAILS OF THE NUMERICAL IMPLEMENTATION
\section{Details of the Numerical Implementation }
\label{app:numerical}
	
The MaxEnt reconstruction of the greybody factors from the BH moment problem has been implemented in
Python. We used SciPy~\cite{Virtanen2020} for the L-BFGS-B minimization and the numerical quadrature,
NumPy for array operations, and Maxima~\cite{maxima} for the symbolic computation of the KdV integrals.

Before the minimization, the moments were rescaled as in Eq.~\eqref{rescaling} for the case without compactification and as in Eq.~\eqref{rescaling_compact} for the case with compactification to improve the numerical
conditioning. The dual potential $\Gamma$~\eqref{dual-potential} and $\Delta$~\eqref{Delta} were then minimized with L-BFGS-B,
using a maximum of 10000 iterations 
%and a tolerance of $10^{-12}$ on the gradient%
, together with a relative tolerance of $10^{-15}$ on the decrease of the objective function (ftol), and a tolerance of $10^{-10}$ on the projected gradient (gtol). For the case without compactification, a
bound $\tilde{\lambda}_{2N}>0$ on the highest-order multiplier was also set.

The integral defining the partition function
$Z$~\eqref{partition-function} was evaluated with adaptive Gauss--Kronrod quadrature, truncated at a
cutoff $\tilde{k}'_{\rm max}$ in the rescaled variable. The cutoff must be large enough for the integrand to be
negligible beyond it, and since this depends on the potential, we took $\tilde{k}'_{\rm max}=100$ for the
P\"oschl--Teller potential (for all $\beta$) and $\tilde{k}'_{\rm max}=4$, $5$ and $7$ for the Regge--Wheeler
potential with $\ell=2$, $5$ and $10$, respectively. Finally, the exponent in the integrand was clipped
to $[-600,600]$ to prevent numerical overflow during the optimization.

%%%%% REFERENCES
%\bibliography{references}

%%%%% DOCUMENT ENDS HERE
\end{document}